\documentclass[letterpaper,12pt]{article}
\usepackage{amssymb,enumerate,float,amsmath}
\usepackage[title]{appendix}

\usepackage{ifpdf}
\ifpdf
\usepackage[pdftex]{graphicx}
\usepackage[pdftex,unicode,implicit]{hyperref}

\hypersetup{
	pdftitle     = {}, 
	pdfkeywords  = {},
	pdfauthor    = {},
	pdfcreator   = {pdf\LaTeXe\ with package \flqq hyperref\frqq},
	pdfproducer  = {pdf\LaTeXe\ with package \flqq hyperref\frqq},
	pdfpagemode  = UseNone,  
	pdffitwindow = true,  
	unicode      = true,
	plainpages   = true,
	colorlinks   = true,  
	citecolor    = black,  
	urlcolor     = blue, 
	linkcolor    = black
}

\else

\usepackage[dvips]{graphicx}

\fi

\makeatletter
\@addtoreset{equation}{section}
\makeatother

\begin{document}
	

\thispagestyle{empty}
	
\begin{center}
{\bf \LARGE Cancellation of divergences in the nonprojectable Ho\v{r}ava theory}
\vspace*{15mm}
		
{\large Jorge Bellor\'{\i}n}$^{1}$,
{\large Claudio B\'orquez}$^{2}$
{\large and Byron Droguett}$^{3}$
\vspace{3ex}
	
{\it Department of Physics, Universidad de Antofagasta, 1240000 Antofagasta, Chile.}
\vspace{3ex}

$^1${\tt jorge.bellorin@uantof.cl,} \hspace{1em}
$^2${\tt cl.borquezg@gmail.com,}
\hspace{1em}
$^3${\tt byron.droguett@ua.cl}
		
\vspace*{15mm}
{\bf Abstract}
\begin{quotation}{ \small \noindent
  We perform an analysis of the ultraviolet divergences of the quantum nonprojectable Ho\v{r}ava gravity. We work the quantum field theory directly in the Hamiltonian formalism provided by the Batalin-Fradkin-Vilkovisky quantization. In this way the second-class constraints can be incorporated to the quantization. A known local gauge-fixing condition leads to a local canonical Lagrangian. Although the canonical fields acquire regular propagators, irregular propagators persist for the field associated to the measure of the second-class constraints. Loops can be formed with the irregular propagators producing potentially dangerous subdivergences. We show that all these loops cancel exactly between them due to a perfect matching between the propagators and vertices of the fields and ghosts forming the loops. The rest of divergences behave similarly to the projectable theory, they can be removed by local counterterms. This result points to the renormalization of the nonprojectable Ho\v{r}ava theory.
}
\end{quotation}
\end{center}

\newpage

\section{Introduction}
The Ho\v{r}ava theory \cite{Horava:2009uw} is a proposal for quantum gravity. It may be formulated as a quantum-field theory with perturbative interactions. In this sense, the theory has the remarkable feature of being power-counting renormalizable and ghost-free.\footnote{Actually, there are regions of the space of coupling constants where ghosts are present; see, for example, \cite{Blas:2009qj}.} Its main characteristic is the breaking of the local Lorentz symmetry, due to the definition of a foliation of spacelike hypersurfaces along a given direction of time with absolute physical meaning. This enables the definition of an anisotropic Lagrangian that is power-counting renormalizable, thanks to the higher order in spatial derivatives. Thus, in this proposal the breaking of the local Lorentz symmetry is the price to pay for obtaining the power-counting renormalizability, simultaneously with the elimination of ghostlike modes, which are known to be present in relativistic higher-curvature theories \cite{Stelle:1976gc}. The underlying gauge symmetry to write the Lagrangian is given by the group of foliation-preserving diffeomorphisms. This symmetry is compatible with two independent formulations of the theory: the projectable and nonprojectable cases. In the projectable case one of the field variables, the lapse function, is restricted to be a function only of time, whereas in the nonprojectable case it can be a general function of time and space. The nonprojectable case has field equations closer to general relativity; hence, it can be more interesting from a physical point of view.

Due to its interest as a possible theory for quantum gravity, several studies have been devoted to analyze the consistency of the theory, both at classical and quantum levels. An essential extension of the Lagrangian of the nonprojectable case was found in Ref.~\cite{Blas:2009qj}, incorporating more terms that are compatible with the foliation-preserving diffeomorphisms. These terms improve the dynamics of the classical theory. The consistency of the classical Hamiltonian formulation for this version has been shown in Refs.~\cite{Kluson:2010nf,Donnelly:2011df,Bellorin:2011ff}. On the projectable side, the Hamiltonian analysis (for the effective action) was carried out in Ref.~\cite{Kobakhidze:2009zr}. 

One of the highlighting aspects of the dynamics of the nonprojectable case is the presence of second-class constraints. Their origin can be regarded qualitatively as a consequence of the reduced symmetry. As it is well known, general relativity has first-class constraints that are the generators of general diffeomorphisms over the spacetime. An analog of the Hamiltonian constraint is still present in the Ho\v{r}ava theory, but it acquires a second-class behavior since there is no generator of a gauge symmetry associated with the transformation of time with space. The other second-class constraint is a formal statement: the vanishing of the conjugate momentum of the lapse function. In the projectable case the symmetry is equally reduced, but the simplification on the lapse function renders the Hamiltonian constraint to be a global (integrated) constraint; hence, effectively there is no local second-class constraint that affects the functional quantum modes.

The main task to do about the self-consistency of the Ho\v{r}ava theory is to study its renormalization beyond the power-counting criterion. In the case of the projectable theory, its renormalizability has been shown in Ref.~\cite{Barvinsky:2015kil}. Despite the similarities, the presence of the second-class constraints on the side of the nonprojectable version constitutes an important technical difference between the two cases. Thus, although the quantization of the projectable case has been focused essentially by following the procedure of gauge theories, this is not so direct in the case of the nonprojectable theory. Other studies concerning the renormalization and renormalization group of the projectable Ho\v{r}ava theory can be found, among others, in Refs.~\cite{Orlando:2009en,Contillo:2013fua,D'Odorico:2014iha,Benedetti:2013pya,Barvinsky:2017kob,Griffin:2017wvh,Barvinsky:2019rwn,Barvinsky:2021ubv}. On the nonprojectable case, an early study about the effect of loops on the coupling to relativistic matter fields was performed in Ref.~\cite{Pospelov:2010mp}. A computation of the heat kernel was focused on in \cite{D'Odorico:2015yaa}.

An essential step in the proof of renormalization of the projectable case consists of obtaining regular propagators for all quantum fields \cite{Barvinsky:2015kil}. The definition of regular propagators was introduced in studies of renormalizability of nonrelativistic gauge theories \cite{Anselmi:2008bq}, and implies that at the ultraviolet they acquire suitable scalings both in frequency and spatial momentum. Once the regularity of all propagators is achieved, the complete renormalization is proved in a similar way as relativistic gauge theories. The gauge-fixing condition chosen for the quantization plays a central role in obtaining the regular propagators for all quantum fields. Remarkably, it turns out that it must be nonlocal. This analysis is based on the Lagrangian formalism under which the classical theory is originally formulated, as usual in the quantization of gauge theories. In Ref.~\cite{Barvinsky:2015kil} it was noticed that a reformulation of the same theory in terms of a Hamiltonian formalism may lead to a local canonical Lagrangian with also regular propagators. This issue is rather nondirect, since the nonlocal gauge-fixing condition is also a noncanonical condition. In Refs.~\cite{Bellorin:2021tkk,Bellorin:2021udn} we showed that the Batalin-Fradkin-Vilkovisky (BFV) quantization may be used in Ho\v{r}ava theory to conciliate these two aspects: the Hamiltonian approach and the  noncanonical gauge-fixing conditions. Previously, two of us focused on the quantization of the nonprojectable case using a canonical gauge, the transverse gauge \cite{Bellorin:2019gsc}. In this case the quantization is performed in the original phase space using the measures for the first- and second-class constraints \cite{Faddeev:1969su,Senjanovic:1976br}. In Ref.~\cite{Bellorin:2021udn} we showed that when the gauge-fixing condition is introduced with a local operator in the BFV quantization of the projectable case, then the integration on momenta of its linearized version reproduces the renormalizable quantum Lagrangian. In this way the connection between the two approaches, BFV and Lagrangian quantization, is established.

Once the BFV quantization of the Ho\v{r}ava theory has been established, in this paper we utilize this scheme of quantization to study loop divergences in the nonprojectable Ho\v{r}ava theory, with the aim of focusing its renormalizability. For definiteness, we take the $2+1$ theory, but we discuss that the approach and the results apply automatically to the $3+1$ theory. Our approach consists of analyzing the quantum theory directly on the Hamiltonian formalism, deriving Feynman rules for all the canonical and auxiliary fields (Lagrange multipliers, ghosts). We want to stay as close as possible to the quantization of the projectable case; hence, we use the previously commented gauge-fixing condition. Indeed, we find that most of the quantum fields get regular propagators.

The BFV formalism introduces an extension of the phase space. In particular, canonical ghost fields, which we call the BFV ghosts, are included. In this way the first-class constraints and the gauge fixing are not imposed in the measure of the path integral. Instead, the second-class constraints are imposed in the standard way as part of the measure. Technically, this measure can be promoted to the quantum canonical Lagrangian with the help of auxiliary fields. Thus, one may develop a Feynman diagrammatic for the theory, including the auxiliary variables. A central aspect in the quantization is that despite the gauge-fixing condition, the auxiliary fields associated with the second-class constraints acquire irregular propagators. Our main focus is on the role of these irregular propagators, since the regular ones lead to well-behaved divergences, like in the projectable case \cite{Barvinsky:2015kil}. Our aim is to characterize the divergences associated to the irregular propagators, in order to study to what extent they can be canceled. This is an important analysis for the renormalizability of the nonprojectable theory, since the qualitative difference with the projectable case is the presence of the second-class constraints, as we have pointed out. The imposition of the second-class constraints is the new ingredient that leads to the irregular propagators.


\section{Quantization in the Hamiltonian formalism}
Here we present a summary of the BFV quantization of the nonprojectable Ho\v{r}ava theory on a foliation of $2+1$ dimensions, with a specific gauge-fixing condition. Further details about the procedure can be found in Refs.~\cite{Bellorin:2021tkk,Bellorin:2021udn}. The BFV quantization of systems with constraints was developed in the papers \cite{Fradkin:1975cq,Batalin:1977pb,Fradkin:1977xi}, starting the study with bosonic systems with first-class constraints only and ending with general systems with bosonic and fermionic constraints that can be of first and second class. This scheme of quantization is based on a Hamiltonian formalism. It allows us to introduce a more general class of gauge-fixing conditions, in particular the noncanonical gauges. 

We start with the definition of the classical theory. Given the foliation with absolute physical meaning, the action of the nonprojectable theory is  
\begin{equation}
 S = \int dt d^2x \sqrt{g} N \left( K_{ij} K^{ij} - \lambda K^2 - \mathcal{V} \right) \,,
\end{equation}
where the independent fields are the Arnowitt-Deser-Misner variables $N(t,\vec{x})$, $N^i(t,\vec{x})$, and $g_{ij}(t,\vec{x})$. The kinetic terms of the Lagrangain are defined in terms of the extrinsic curvature tensor,
\begin{equation}
 K_{ij} = \frac{1}{2N} \left( \dot{g}_{ij} - 2 \nabla_{(i} N_{j)} \right) \,,
 \qquad
 K = g^{ij} K_{ij} \,.
\end{equation}
$\mathcal{V}$ is usually called the potential. It contains the higher-order spatial derivatives, whose order is identified with the parameter $z$, such that the higher-order terms are of $2z$ order in spatial derivatives. Thus, $z$ parametrizes the degree of anisotropy of the theory, since it can be $z > 1$. In this anisotropic framework, time and space are declared to have dimensions
\begin{equation}
	[t]=-z,\qquad [x^i]=-1 \,.
\end{equation}
To have an anisotropic and power-counting renormalizable theory in $2+1$ dimensions, $\mathcal{V}$ must be of order $z = 2$. In this case, the complete potential is \cite{Colombo:2014lta}
\begin{eqnarray}
\mathcal{V} &=&
-\beta R-\alpha a^{2}+\alpha_{1}R^{2}+\alpha_{2}a^{4}+\alpha_{3}R a^{2}+\alpha_{4}a^{2}\nabla_{i}a^{i}
\nonumber \\ &&
+\alpha_{5} R\nabla_{i}a^{i} 
+\alpha_{6}\nabla^{i}a^{j}\nabla_{i}a_{j}+ \alpha_{7}(\nabla_{i}a^{i})^{2} \,,
\label{potencial}
\end{eqnarray}
where 
\begin{equation}
a_i = \frac{\partial_i N}{N} 
\end{equation}
is a vector that enters in the Lagrangian in a way compatible with the gauge symmetry \cite{Blas:2009qj}. $\nabla_i$ and $R$ are the covariant derivative and the Ricci scalar of the spatial metric $g_{ij}$. The coupling constants of the theory are $\lambda$, $\beta$, $\alpha$, $\alpha_1$,...,$\alpha_7$. We use shorthand for frequent combinations of these coupling constants,
\begin{equation}
 \sigma = \frac{1-\lambda}{1-2\lambda} \,,
 \quad
 \bar{\sigma} = \frac{\lambda}{1-2\lambda} \,,
 \quad
 \alpha_{67} = \alpha_6 + \alpha_7 \,.
\end{equation} 

After the Legendre transformation, the primary classical Hamiltonian results
\begin{equation}
	H_0 =
	\int d^{2}x
	\sqrt{g} N \left( \frac{\pi^{ij}\pi_{ij}}{g} 
	+ \bar{\sigma} \frac{\pi^{2}}{g} + \mathcal{V} \right) \,.
	\label{H0}
\end{equation}
The canonically conjugate pairs are $(g_{ij},\pi^{ij})$ and $(N,P_N)$. The unique first-class constraint the theory has is 
\begin{equation}
	\mathcal{H}_i =
	- 2 g_{ij} \nabla_k \pi^{kj} = 0 \,.
	\label{momentumconts}
\end{equation}
The second-class constraints are 
\begin{eqnarray}
	\theta_{1} &=&  
	\sqrt{g} N \left( \frac{\pi^{ij}\pi_{ij}}{g} 
	+ \bar{\sigma} \frac{\pi^{2}}{g} + \mathcal{V} \right) 
	+ \sqrt{g} \Big( 
	  2\alpha\nabla_{i}(Na^{i})
	- 4\alpha_{2}\nabla_{i}(Na^{2}a^{i})
	- 2\alpha_{3}\nabla_{i}(NRa^{i})
	\nonumber\\ &&
	+ \alpha_{4} \left( \nabla^{2}(Na^{2}) 
	-2\nabla_{i}( N a^{i} \nabla_{j} a^{j} ) \right)
	+ \alpha_{5}\nabla^{2}(NR)
	+ 2\alpha_{6}\nabla^{i}\nabla^{j}(N\nabla_{j}a_{i})
	\nonumber \\ &&
	+ 2\alpha_{7}\nabla^{2}(N\nabla_{i}a^{i})
	\Big) 
    = 0 \,,
	\label{theta2}
	\\ 
	\theta_{2} &=& P_{N}=0 \,.
    \label{theta1}
\end{eqnarray}
Following Dirac's procedure, constraint $\theta_2$ can be regarded as a primary constraint since in the Lagrangian there is no time derivative of $N$, whereas $\theta_1$ is the consequence of preserving $\theta_2$. An important feature is that the primary Hamiltonian (\ref{H0}) can be written as the integral of one of the second-class constraints,
\begin{equation}
	H_{0} = \int d^{2}x \,\theta_{1} \,.
	\label{primaryhamiltonian}
\end{equation}
In the classical canonical formulation the shift vector $N^i$ is regarded as the Lagrange multiplier associated with the first-class constraint (\ref{momentumconts}).

The BFV quantization of the nonprojectable Ho\v{r}ava theory requires to extend the phase space by adding the canonical pairs $(N^i,\pi_i)$, $(C^i,\bar{\mathcal{P}}_i)$, and $(\bar{C}_i,\mathcal{P}^i)$. The last two pairs are the BFV ghosts. The definition of the BFV path integral, under a given gauge-fixing condition, is
\begin{equation}
Z =
\int \mathcal{D}V e^{iS},
\label{Z}
\end{equation}
where the measure and the action are given, respectively, by
\begin{eqnarray}
&&
\mathcal{D}V = 
\mathcal{D} g_{ij} \mathcal{D}\pi^{ij} \mathcal{D}N \mathcal{D}P_{N}\mathcal{D} N^{k}\mathcal{D}\pi_{k}
\mathcal{D} C^i \mathcal{D} \bar{\mathcal{P}}_i
\mathcal{D} \bar{C}_i \mathcal{D} \mathcal{P}^i
\times \delta(\theta_{1}) \delta(\theta_{2}) \sqrt{\det\{\theta_{p},\theta_{q}\}} \,, \;\;\;
\label{medida}
\\ &&
S=
\int dt d^{2}x \left( 
  \pi^{ij} \dot{g_{ij}} 
+ P_N \dot{N} 
+ \pi_i \dot{N}^i 
+ \bar{\mathcal{P}}_{i} \dot{C}^{i} 
+ \mathcal{P}^{i} \dot{\bar{C}}_{i} 
- \mathcal{H}_{\Psi}
\right) \,.
\label{scan} 
\end{eqnarray}
The quantum gauge-fixed Hamiltonian density is
\begin{equation}
 \mathcal{H}_{\Psi} =
 \mathcal{H}_0 + \{\Psi,\Omega\}_{\text{D}} \,,
 \label{bfvhamiltonian}
\end{equation}
where $\Omega$ is the generator of the Becchi-Rouet-Stora-Tyutin (BRST) symmetry, $\Psi$ is a gauge-fixing fermionic function and $\{\,,\}_{\text{D}}$ indicates Dirac brackets. A theorem developed in the original formulation of the BFV quantization \cite{Fradkin:1975cq,Batalin:1977pb,Fradkin:1977xi} ensures that the above path integral is independent of the choice of the function $\Psi$.  The BRST generator for the nonprojectable Ho\v{r}ava theory becomes
\begin{equation}
\Omega = 
\mathcal{H}_{i} C^{i} + \pi_{i} \mathcal{P}^{i} 
- \frac{1}{2} U^{k}_{ij} C^{i} C^{j} \bar{\mathcal{P}}_{k} \,,
\label{Omega}
\end{equation}
where the $U_{ij}^k$ are taken from the algebra of spatial diffeomorphisms (see \cite{Bellorin:2021udn}),
\begin{equation}
  \{ \mathcal{H}_i \,, \mathcal{H}_j \} =
  U_{ij}^k \mathcal{H}_k \,.
\end{equation}
The factor $\delta(\theta_{1}) \delta(\theta_{2}) \sqrt{\det\{\theta_{p},\theta_{q}\}}$ in (\ref{medida}) is the measure associated with the second-class constraints. In the present theory, due to the algebra of constraints, it turns out that
\begin{equation}
\sqrt{\det\{\theta_{p},\theta_{q}\}}=
\det\{\theta_{1},\theta_{2}\} 
\,.
\label{sprtbra}
\end{equation}
The variable $P_N$ may be integrated directly by using the delta $\delta(\theta_2)$ in the measure, such that $P_N$ is eliminated from the Lagrangian (the bracket $\{ \theta_1 \,, \theta_2 \}$ in (\ref{sprtbra}) does not depend on $P_N$). The other delta, $\delta(\theta_1)$, may be promoted to the quantum Lagrangian by means of a Lagrange multiplier,
\begin{equation}
\delta( \theta_1 ) =
\int \mathcal{D} \mathcal{A} \exp\left( i
\int dt d^2x \mathcal{A} \, \theta_1  \right) \,.
\end{equation}
We may also promote the factor (\ref{sprtbra}) to the quantum Lagrangian by means of a pair of fermionic ghosts, $\eta,\bar{\eta}$,
\begin{eqnarray}
\det \{ \theta_1 \,, \theta_2 \} &=& 
\int \mathcal{D}\bar{\eta} \, \mathcal{D} \eta
\exp\left( i \int dt d^2x\, 
\bar{\eta} \{ \theta_1 \,, \theta_2 \} \eta \right) \,.
\label{gostsecond}
\end{eqnarray}
Thus, $\mathcal{A}$ and $\eta,\bar{\eta}$ enter as additional fields in the quantum theory. The scalings of the quantum fields are
\begin{equation}
\begin{array}{lll}
& [g_{ij}]=0 \,,   & [\pi^{ij}] = d = 2 \,, \\[1ex]
& [N]=0 \,,   &  \\[1ex]
& [N^i]=z-1=1 \,,  & [\pi_i]=1+d-z = 1 \,, \\[1ex]
& [C^i] = [\bar{C}_i] = (d-z)/2 = 0 \,, \qquad
& [\mathcal{P}^i] = [\bar{\mathcal{P}}_i] = (d+z)/2 =2 \,, \\[1ex]
& [\mathcal{A}]=z-d=0, & [\bar{\eta}]=[\eta]=(z-d)/2 = 0 \,.
\end{array}
\end{equation}

The aim in the BFV formulation is to use a gauge-fixing condition and an associated fermionic function of the general form
\begin{eqnarray}
 &&
 \dot{N}^i - \chi^i = 0\,,
 \label{relativisticgaugephi}
 \\ &&
 \Psi =
 \bar{\mathcal{P}}_{i} N^{i} + \bar{C}_{i} \chi^{i} \,.
 \label{relativisticgaugepsi}
\end{eqnarray} 
Equation (\ref{relativisticgaugephi}) expresses the noncanonical nature of the gauge fixing: it depends on time derivatives. This kind of gauge-fixing condition was originally introduced in the BFV quantization to make compatible the Hamiltonian formalism with the relativistic gauges (especially for unitarity). It turns out that, since the functional part $\chi^i$ is left free, one may use this form of gauge fixing both for relativistic and nonrelativistic gauges. By using (\ref{Omega}) and (\ref{relativisticgaugepsi}), and assuming that $\chi^i$ does not depend on the BFV ghosts, the gauge-fixed Hamiltonian (\ref{bfvhamiltonian}) takes the form
\begin{equation}
\begin{split}
\mathcal{H}_{\Psi} = \, &
  \mathcal{H}_{0} 
+ \mathcal{H}_iN^{i}
+ \bar{\mathcal{P}}_i \mathcal{P}^{i}
- \bar{\mathcal{P}}_{i} \left(N^{j}\partial_{j} C^{i} +  
       N^{i} \partial_{j} C^{j} \right)
+\pi_i\chi^{i}
\\ &
+ \bar{C}_i \{\chi^{i} \,, \mathcal{H}_j\}_{\mathrm{D}} C^{j}
+ \bar{C}_i \dfrac{ \delta\chi^{i}}{\delta N^{j}} \mathcal{P}^{j} \,.
\label{genhamiltonian}
\end{split}
\end{equation}

Now we specify the gauge-fixing condition. To this end, we move to perturbative variables, defined by
\begin{equation}
 g_{ij} = \delta_{ij} + h_{ij} \,,
 \quad
 \pi^{ij} = p^{ij} \,,
 \quad
 N = 1 + n \,,
\quad
N^i = n^i \,.
\end{equation}
We denote traces of tensors by $h = h_{kk}$ and $p = p^{kk}$. On the rest of quantum fields we keep the original notation, considering them as perturbative variables. We introduce the gauge-fixing condition in the form
\begin{equation}
	\chi^{i}= 
	\mathfrak{D}^{ij}\pi_{j} + \Gamma^{i} \,,
	\label{chi}
\end{equation}
where
\begin{eqnarray}
  &&
	\mathfrak{D}^{ij} = \delta_{ij} \Delta + \kappa \partial_{ij}  \,,
	\label{D}
 \\ &&
 \Gamma^{i} =
 - 2 \Delta\partial_jh_{ij}
 + 2 \lambda (1+\kappa) \Delta\partial_{i}h
 - 2\kappa \partial_{ijk} h_{jk} \,.
 \label{gammaproj}
\end{eqnarray}
$\kappa$ is an arbitrary constant. We use the shorthand $\partial_{ij\cdots k} = \partial_i \partial_j \cdots \partial_k$. It is known \cite{Bellorin:2021udn} that by using this local gauge in the projectable theory, after the integration on momenta, the nonlocal quantum Lagrangian of Ref.~\cite{Barvinsky:2015kil} is recovered. The coefficients in the gauge fixing are chosen to simplify the resulting propagators of the quantum fields. Other frequent combinations of constants that arise are
\begin{equation}
\bar{\kappa} = \kappa + 1 \,, 
\quad
\rho_1 = 2 (1-\lambda)\bar{\kappa} \,,
\quad
\rho_2 = 4 \alpha_{1}-\frac{\alpha_{5}^{2}}{\alpha_{67}} \,.
\end{equation}
Under the gauge-fixing condition (\ref{chi}), the path integral takes the form
\begin{equation}
\begin{split}
Z =& 
\int \mathcal{D} h_{ij} \mathcal{D}p^{ij} \mathcal{D}n \mathcal{D}n^{i} \mathcal{D}\pi_{i} \mathcal{D} C^i \mathcal{D}\bar{\mathcal{P}}_i \mathcal{D}\bar{C}_i \mathcal{D}\mathcal{P}^i 
\mathcal{D}\mathcal{A}
\mathcal{D}\bar{\eta}
\mathcal{D}\eta 
\\ &
\exp\left[ i \int dt d^{2}x 
\Big(p^{ij}\dot{h}_{ij} +\pi_{i}\dot{n}^{i}
+\bar{\mathcal{P}}_i \dot{C}^i + \bar{C}_i \dot{\mathcal{P}}^i
- \mathcal{H}_{0} 
- \mathcal{H}_in^{i}
- \bar{\mathcal{P}}_i \mathcal{P}^{i}
+ \bar{\mathcal{P}}_{i} ( n^{j} \partial_{j} C^{i} 
\right. 
\\ &   
+ n^{i}\partial_{j} C^{j} )
- \pi_i \mathfrak{D}^{ij} \pi_{j}
+ 2 \pi_i ( 
      \Delta\partial_jh_{ij}
	- \lambda \bar{\kappa} \Delta\partial_{i}h
	+ \kappa \partial_{ijk} h_{jk} )
\\ &
- \bar{C}_i\{\Gamma^{i} \,, \mathcal{H}_j\} C^{j}
\left.
+ \mathcal{A}\theta_{1}
+ \bar{\eta} \{ \theta_1 \,, \theta_2 \} \eta
\Big)\right]
\,.
\end{split}
\label{pathintegralguagefix}
\end{equation}
Some terms in this Lagrangian must be expanded in perturbations. Here we show them at quadratic order, in appendix A we show their expressions at cubic order,
\begin{equation}
\begin{split}
&
 \mathcal{H}_{0} =
   p^{ij} p^{ij} + \bar{\sigma} p^{2}
 + \beta \left( 
   \frac{1}{4} h\Delta h 
 - \frac{1}{4} h_{ij} \Delta  h_{ij}
 + \frac{1}{2} h_{ij} \partial_{ik} h_{jk}
 + n \Delta h 
 - n\partial_{ij} h_{ij}
 - \frac{1}{2} h \partial_{ij} h_{ij} \right)
 \\ &
  + \alpha n\Delta  n
  + \alpha_{1} \Big( h \Delta^2 h + h_{ij} \partial_{ijkl} h_{kl}
  - 2 h\Delta \partial_{ij} h_{ij} \Big)
  - \alpha_{5} n \left( 
     \Delta^2 h
  -  \Delta \partial_{ij} h_{ij} \right)
  + \alpha_{67} n \Delta^2  n \,,
  \\ &
  \mathcal{H}_i n^{i} =
  - 2 n^i \partial_j p^{ij} \,,
  \\ &
  \bar{C}_i  \{\Gamma^{i} \,, \mathcal{H}_j \} C^{j} =
  - 2 \bar{C}^{i} \left( \delta_{ij} \Delta^2 + ( \rho_1 - 1 )\Delta \partial_{ij} \right) C_j \,, 
  \\ &
  \bar{\eta} \{ \theta_1 \,, \theta_2 \} \eta =	
  2 \alpha_{67} \bar{\eta} \Delta^2 \eta \,,
  \\ &
  \mathcal{A} \theta_1^{(1)} =
	\mathcal{A} \left( 
	  \beta(\Delta h - \partial_{ij} h_{ij} )
	+ 2\alpha\Delta n
	- \alpha_5\Delta ( \Delta h - \partial_{ij} h_{ij} )
	+ 2\alpha_{67}\Delta^2 n 
	\right) \,.
\end{split}
\end{equation}

This quantum formulation of the nonprojectable Ho\v{r}ava theory has a BRST symmetry when the theory is evaluated on the surface of the phase space defined by the second-class constraints. By definition, the measure of the second-class constraints becomes $1$ evaluated on this surface. Therefore, the field variables $\mathcal{A}$ and $\eta,\bar{\eta}$ are not relevant for this symmetry since they are associated precisely with this measure (indeed, they are not canonical variables). 

\section{Loops in diagrams}
We may define Feynman rules in the framework of the BFV formalism. The first step is to obtain the propagators of the quantum fields (canonical fields, canonical BFV ghosts and the fields $\mathcal{A}$ and $\eta,\bar{\eta}$). A clean way to achieve this is by performing a transverse-longitudinal decomposition on all vector and tensor fields in the second-order  Lagrangian, finding the propagators by inversion, and then reconstructing the vector and tensor propagators \cite{Barvinsky:2015kil}. We use the two-dimensional transverse-longitudinal decomposition for tensors,
\begin{equation}
\begin{split}
  & 
  h_{ij}=
  \Big(\delta_{ij} - \frac{\partial_{ij}}{\Delta }\Big)h^T
  + \partial_{(i}h^L_{j)}
  + \frac{\partial_{ij}}{\Delta} h^L \,,
  \\ & 
  p^{ij}=
  \Big(\delta_{ij} - \frac{\partial_{ij}}{\Delta }\Big)p^T
  + \partial_{(i}p^L_{j)}
  + \frac{\partial_{ij}}{\Delta} p^L \,,
  \\ &
  \partial_ih_i^L=\partial_ip_i^L=0 \,,
\end{split}
\end{equation} 
and vectors:
\begin{equation}
 n^i=n_i^T+\partial_in^L,\qquad \pi_i=\pi_i^T+\partial_i\pi^L,\qquad \partial_in_i^T=\partial_i\pi_i^T=0 \,.
\label{vectors}
\end{equation}
The four BFV ghost fields are also decomposed as in (\ref{vectors}). From now on we focus on the high-energy regime; hence, we send the low-order coupling constants $\beta$ and $\alpha$ to zero. The canonical quantum Lagrangian at second order becomes
\begin{equation}
\begin{split} 
\mathcal{L}^{(2)} = &
\bar{\mathcal{P}}_{i}^{T}\dot{C}^{T}_i
- \bar{\mathcal{P}}^{L}\Delta \dot{C}^{L}
+ \bar{C}_{i}^{T}\dot{\mathcal{P}}^{T}_i
- \bar{C}^{L}\Delta\dot{\mathcal{P}}^{L}
- \bar{\mathcal{P}}_{k}^{T}\mathcal{P}^{T}_k
+ \bar{\mathcal{P}}^{L}\Delta\mathcal{P}^{L}
+ 2\bar{C}_{k}^{T}\Delta^{2}C^{T}_k
\\ &
- 2 \rho_1 \bar{C}^{L}\Delta^{3}C^{L}
+ p^{T}\left(\dot{h}^{T}- 2 \bar{\sigma} p^{L}\right)
- \frac{1}{2} p^{L}_i \Delta\dot{h}_{i}^{L}
- \sigma \left(p^{T}\right)^{2}
+ \frac{1}{2}p^{L}_i\Delta p^{L}_i
\\ &
+ p^{L}\dot{h}^{L}
- \sigma \left(p^{L}\right)^{2}
+ p^{L}_i\Delta n^{T}_i
- 2p^{L}\Delta n^{L}
- \alpha_{1}h^{T}\Delta^{2}h^{T}
+ \alpha_{5}n\Delta^{2}h^{T}
\\ &
- \alpha_{67}n\Delta^{2}n
+ \pi_{k}^{T}\dot{n}^{T}_k
- \pi^{L}\Delta\dot{n}^{L}
- \pi_{k}^{T}\Delta\pi_{k}^{T}
+ \bar{\kappa}\pi^{L}\Delta^{2}\pi^{L}
+ \pi_{k}^{T}\Delta^{2}h_k^{L}
\\ &
+ 2\lambda\bar{\kappa}\pi^{L}\Delta^{2}h^{T}
- \rho_1 \pi^{L}\Delta^{2}h^{L}
+ \mathcal{A}\left(-\alpha_{5}\Delta^{2}h^{T}
+ 2 \alpha_{67} \Delta^{2}n\right)
+ 2 \alpha_{67} \bar{\eta}\Delta^{2}\eta \,.
\end{split}
\label{secondorderlagrangian}
\end{equation}

We present all the nonzero propagators derived from the quantum Lagrangian (\ref{secondorderlagrangian}).\footnote{In the complete canonical Lagrangian (\ref{pathintegralguagefix}) there are no vertices for the canonical momenta $\pi_i$ and $\mathcal{P}_i$, hence these fields do not arise in connected diagrams. This could be changed by a different gauge-fixing condition, but this is not relevant for the present study.} In Fourier space, after a Wick rotation, they are
\begin{equation}
\begin{split}
\langle \bar{\mathcal{P}}_i\mathcal{P}^{j}\rangle =&
 4 k^4 \left( P_{ij}\mathcal{T}_{3} 
 + 2 \rho_1 \hat{k}_{i} \hat{k}_{j} \mathcal{T}_{2} \right)
\,,
\\
\langle \mathcal{P}_i\bar{C}^{j}\rangle =& 
- \langle \bar{\mathcal{P}}_iC^{j}\rangle =
\omega \mathcal{S}_{ij} \,,
\quad
\langle \bar{C}_iC^{j}\rangle = - \mathcal{S}_{ij}  \,,
\\
\langle p^{ij}p^{kl} \rangle =&
 \rho_2 k^{4} P_{ij} P_{kl}\mathcal{T}_{1} \,,
\\
\langle h_{ij} h_{kl} \rangle =&
 2 Q_{ijkl} \left( 1 - 2 k^4 \mathcal{T}_{3}\right)\mathcal{T}_{3}
 + 4 \left( {\sigma} P_{ij} P_{kl}
 + \bar{\sigma} \hat{k}_i \hat{k}_j P_{kl} 
 + \bar{\sigma} \hat{k}_k \hat{k}_l P_{ij} 
 \right) \mathcal{T}_{1}
 + 4 \sigma Q \hat{k}_{i} \hat{k}_{j} \hat{k}_{k} \hat{k}_{l} \mathcal{T}_{1} \mathcal{T}_{2}^{2} \,,
\\
\langle h_{ij} p^{kl} \rangle =&
 \omega \left[ Q_{ijkl} \mathcal{T}_{3}
 + 2 P_{ij} P_{kl}\mathcal{T}_{1}
 + 2 \bar{\sigma} \left( \rho_2 + 2 \rho_1 - 4 \lambda \bar{\kappa} \right) k^2 k_{i}k_{j} P_{kl} \mathcal{T}_{1}\mathcal{T}_{2}
 + 2 \hat{k}_{i} \hat{k}_{j} \hat{k}_{k} \hat{k}_{l} \mathcal{T}_{2} \right] \,,
\\
\langle n^{i} h_{jk} \rangle =&
  16 i \omega k^3 \left( P_{i(j} \hat{k}_{k)} \mathcal{T}_{3}^{2}
  + \bar{\kappa} \hat{k}_i \hat{k}_j \hat{k}_k \mathcal{T}_{2}^{2} \right) \,,
\\
\langle n^{i} p^{jk} \rangle =&
  2 i k^3 \left[ 2 P_{i(j} \hat{k}_{k)} \mathcal{T}_{3}
  + \hat{k}_{i} \left( \rho_1 \hat{k}_j \hat{k}_{k}
  - 2 \lambda\bar{\kappa} P_{jk} \right) \mathcal{T}_{2} 
   \right] \,,
\\
\langle n^{i} n^{j} \rangle =&
 - 4 k^2 P_{ij} \mathcal{T}_{3}(1-4k^{4}\mathcal{T}_{3})
 - 4 \bar{\kappa} \omega^{2} \left[ \omega^{2} - \left( \rho_1 + \sigma \rho_2 ( 2 \rho_1 + 1 ) \right) k^4 \right] k_{i}k_{j} 
 \mathcal{T}_{1} \mathcal{T}_{2}^{2} \,,
\\ 
 \langle h_{ij} n \rangle =&
 \frac{2\alpha_{5}}{\alpha_{67}} \left( \sigma P_{ij}
 + \bar{\sigma} \hat{k}_i \hat{k}_j \right)\mathcal{T}_{1} \,,
\quad
 \langle p^{ij} n \rangle =
  \frac{\alpha_{5}}{\alpha_{67}}\omega P_{ij}\mathcal{T}_{1} \,,
\quad
\langle n^{i}\pi_{j} \rangle =
  \omega \mathcal{S}_{ij}  \,,
\\
\langle \pi_i h_{jk} \rangle =&
 - 4 i \left( P_{i(j} k_{k)} \mathcal{T}_{2}
 + k_{i} \hat{k}_{j} \hat{k}_{k}\mathcal{T}_{3} \right) \,,
\quad
\langle nn \rangle =
     \frac{ \alpha_{5}^2 \sigma }{ \alpha_{67}^2 } \mathcal{T}_{1} \,,
\\
\langle \mathcal{A}\mathcal{A} \rangle =&
\langle \mathcal{A} n \rangle = 
  \langle \eta \bar{\eta} \rangle = 
  \frac{1}{ \alpha_{67} k^{4}} \,,
\end{split}
\label{propagators}
\end{equation}
where
\begin{equation}
\begin{split}
&
\hat{k}_i = \frac{ k_i }{k} \,,
\quad
P_{ij} = \delta_{ij} - \hat{k}_{i} \hat{k}_{j} \,,
\\ &
Q_{ijkl} =
  \hat{k}_i \hat{k}_k P_{jl} + \hat{k}_j \hat{k}_k P_{il} 
+ \hat{k}_i \hat{k}_l P_{jk} + \hat{k}_j \hat{k}_l P_{ik} \,,
\\ &
Q =
\omega^{4} 
+ \left( 12\lambda\bar{\kappa}
- 2 (1+2\lambda) \rho_1 - \rho_2 \right) 
\frac{\omega^{2}k^{4}}{1-\lambda} 
+ 4 \bar{\kappa} \left( \rho_2 +4\lambda^{2}\bar{\kappa} \right) k^8  \,,
\\ &
\mathcal{T}_{1} =
\left(\omega^2-\rho_2 {\sigma}k^4\right)^{-1} \,,
\quad
\mathcal{T}_{2} = 
\left(\omega^{2} + 2 \rho_1 k^4 \right)^{-1},
\\ &
\mathcal{T}_{3} =
\left(\omega^{2}+2k^{4}\right)^{-1},
\quad
\mathcal{S}_{ij} =
2 P_{ij}\mathcal{T}_{3} + 2 \hat{k}_{i} \hat{k}_{j} \mathcal{T}_{2} \,.
\end{split}
\end{equation}

The definition of regular propagators was introduced in Ref.~\cite{Anselmi:2008bq} to study the renormalizability of nonrelativistic gauge theories. This was used in the proof of renormalization of the projectable Ho\v{r}ava theory \cite{Barvinsky:2015kil}, whose definition we take here. Consider a propagator between two fields that have scaling dimensions $r_1$ and $r_2$. It is regular if it is given by the sum of terms of the form
\begin{equation}
 \frac{ P(\omega,k^i) }{ D(\omega,k^i) } \,,
\end{equation}
where $D$ is the product
\begin{equation}
 D = 
 \prod_{m=1}^{M} ( A_m \omega^2 + B_m k^{2d} + \cdots ) \,, 
 \quad
 A_m \,, B_m > 0 \,,
 \label{regular}
\end{equation}
and $P(\omega,k^i)$ is a polynomial of maximal scaling degree less than or equal to $r_1 + r_2 + 2(M-1)d$. All the propagators in the list (\ref{propagators}) satisfy this condition of regularity,\footnote{Actually, the condition of regular propagator in (\ref{regular}) requires the coefficients in the denominators to be strictly positive. In the present theory this is satisfied by the regular propagators if $\rho_1 > 0$ and $\sigma \rho_2 < 0$.} except for the three propagators $\langle \mathcal{A} \mathcal{A} \rangle$, $\langle \mathcal{A} n \rangle$, and $\langle \bar{\eta}\eta \rangle$, which are independent of $\omega$. The three irregular propagators involve the variables $\mathcal{A},\eta,\bar{\eta}$, and these are fields associated with the measure of the second-class constraints, as we have indicated.

\paragraph{Superficial degree of divergence}
To define the superficial degree of divergence $D_{\text{div}}$ of each diagram we consider the different scaling between space and time. In the $2+1$-dimensional case we have
\begin{equation}
k^i \rightarrow b k^i \,,
\qquad 
\omega\rightarrow b^{2}\omega \,.
\end{equation}
Given a diagram with $L$ loops, its superficial degree of divergence is given by
\begin{equation}
 D_{\text{div}} = 
 4 L + \sum\limits_r w_r I_r + \sum\limits_s d_s V_s - X \,,
\end{equation}
where $I_r$ and $V_s$ denote the number of propagators and vertices of type $r$ and $s$, respectively, $w_r$ is the scaling for each propagator at the ultraviolet, and $d_s$ is the number of spatial derivatives of each type of vertex. The factor $X$ counts the number of spatial derivatives that may exist in the external legs of the diagram. We remark that the BFV quantization is based on a Hamiltonian formalism; as a consequence there are no time derivatives in the vertices extracted from the canonical Lagrangian. Hence, we do not need to consider time derivatives on the external legs. By inspection of the propagators (\ref{propagators}) and vertices in (\ref{pathintegralguagefix}), and using the topological identity $L - 1 = \sum I_r - \sum V_s$, we obtain\footnote{We use the following notation: $I_{\Psi_1 \Psi_2}$ is the number of internal lines with the propagator $\langle \Psi_1 \Psi_2 \rangle$, $V_{\Psi_1\Psi_2\cdots}$ is the number or vertices with legs $\Psi_1,\Psi_2,\cdots$. Fields are represented by their own symbols omitting indices, except for the index $N$ that stands for the field $n^i$. Square brackets indicate that the vertex has multiple legs for that field.}
\begin{equation}
\begin{split}
D_{\text{div}} =&
4 + 4 I_{pp} + 2 I_{pn} + 2 I_{NN} + 2 I_{hp} + I_{hN} + 3 I_{pN}
+ 2 I_{\bar{\mathcal{P}}C}
\\ &
- 4 V_{\mathcal{A}pp} - 4 V_{pp [n]} - 4V_{pp[h]} - 3V_{Np[h]} 
- 3V_{\bar{\mathcal{P}}CN}-X \,.
\end{split}
\label{sdd}
\end{equation}
For a given field, we can match the number of internal and external lines that contain it with the number of vertices that have legs for it. By doing so for some of the fields, we obtain the identities
\begin{equation}
\begin{split}
&
2 I_{pp} + I_{pn} + I_{hp} + I_{pN} + E_{p} =
2 V_{\mathcal{A}pp} + V_{Np[h]} + 2V_{pp[n]} + 2V_{pp[h]}, 
\\ &
2 I_{NN} + I_{Nh} + I_{pN} + E_{N} =
V_{Np[h]} + V_{\bar{\mathcal{P}}CN} \,,
\\ &
I_{\bar{\mathcal{P}}\mathcal{P}} + I_{\bar{\mathcal{P}}C} =
V_{\bar{\mathcal{P}}CN} \,.
\end{split}
\end{equation}
Finally, by substituting these relations in (\ref{sdd}), we obtain
\begin{eqnarray}
D_{\text{div}}=
4 - E_{N} - 2E_{p} - X.
\end{eqnarray}
Therefore, the diagrams with maximal superficial degree of divergence are those with no external legs for the $N^i$ and $p_{ij}$ fields, and carrying no spatial derivatives on their external legs. These diagrams have a superficial degree of divergence of 4, which is the order of the bare Lagrangian of the $2+1$ nonprojectable Ho\v{r}ava theory. This result is in agreement with the original power-counting renormalizability of the theory.

We comment that in the list of propagators (\ref{propagators}) there are several scalings on $\omega$, in particular the irregular propagators which do not depend on $\omega$. This is the critical issue for the divergences in loops. Since the theory is nonrelativistic, it makes sense to analyze the degree of divergence of the integration on $\omega$ separately, which we denote by $D_{\text{div}}(\omega)$. For a given diagram, it is given by
\begin{equation}
  D_{\text{div}}(\omega) = 2L + \sum\limits_r w_r(\omega) I_r \,,
\end{equation} 
where $w_r(\omega)$ denotes the scaling that each propagator has in $\omega$. By inspection on the propagators in (\ref{propagators}), we find
\begin{equation}
\begin{split}
D_{\text{div}}(\omega) =&
  2 
+ 2 ( I_{\mathcal{A}\mathcal{A}} + I_{\mathcal{A}n} + I_{\bar{\eta}\eta} )
- 2 ( I_{hh} + I_{nn} + I_{pp} + I_{hn} + I_{NN} 
\\ &
+ 2 I_{hN} + I_{pN} + I_{\bar{C}C} ) 
- 2 \sum_s V_s \,.
\end{split}
\end{equation}
This confirms that irregular propagators are the only ingredients that increase the divergence in the $\omega$ direction.

\paragraph{Unavoidable irregular propagators}
We have found irregular propagators for the fields $\mathcal{A}$ and $\eta,\bar{\eta}$. Now we discuss how the irregularity of these propagators is an unavoidable consequence of the structure of the second-class constraints, which in particular affects the variable $n$.

We denote all fields of the theory by $\Psi_A$. The second-order Lagrangian (\ref{secondorderlagrangian}) can be written in matrix form,
\begin{equation}
\mathcal{L}^{(2)} =
\frac{1}{2} \Psi_A M_{AB} \Psi_B \,,
\quad
M_{AB} =
\frac{ \delta^2 \mathcal{L}^{(2)} }{ \delta \Psi_A \delta \Psi_B } \,.
\end{equation}
Each entry of propagators is given by $\langle \Psi_A \Psi_B \rangle = [M^{-1}]_{AB}$. Let us start with the propagators involving $\mathcal{A}$. We denote by $\hat{\Psi}_{\hat{A}}$ the subset of $\Psi_A$ that excludes $\mathcal{A}$. The key point is the fact that  constraint $\theta_1$ is equivalent to the derivative
\begin{equation}
 \theta_1 = \frac{ \delta \mathcal{H}_0 }{\delta n} \,,
\end{equation}
along with the dependence that the canonical Lagrangian has on $n$. Indeed, all the dependence that the quadratic canonical Lagrangian (\ref{secondorderlagrangian}) has on the fields $n$ and $\mathcal{A}$ comes from the structure 
\begin{equation}
\mathcal{L} =
  \mathcal{H}_0^{(2)}
+ \mathcal{A} \frac{\delta \mathcal{H}^{(2)}_0}{\delta n} 
\,.
\label{nAterms}
\end{equation}
From this it is straightforward to obtain the relations
\begin{eqnarray}
&&
\frac{ \delta^2 \mathcal{L} }{ \delta \mathcal{A} \delta \mathcal{A} }
= 0 \,,
\label{dAdA}
\\ &&
\frac{ \delta^2 \mathcal{L} }{ \delta \hat{\Psi}_{\hat{A}} \delta \mathcal{A} }
=
\frac{ \delta^2 \mathcal{L} }{ \delta \hat{\Psi}_{\hat{A}} \delta n } \,.
\label{dpsidA}
\end{eqnarray}	
The determinant of $M$ is given by
\begin{equation}
\det M =
\epsilon_{A B C \cdots D}
\left( \frac{ \delta^2 \mathcal{L} }{ \delta \Psi_A \delta n } \right)
\left( \frac{ \delta^2 \mathcal{L} }{ \delta \Psi_B \delta \mathcal{A} } \right)
W_{C \cdots D} \,,
\label{detM}
\end{equation}
where the sector
\begin{equation}
W_{C \cdots D} =
\left( \frac{ \delta^2 \mathcal{L} }{ \delta \Psi_C \delta h^T } \right) 
\cdots
\left( \frac{ \delta^2 \mathcal{L} }{ \delta \Psi_D \delta \eta } \right)
\end{equation}
covers all columns of $M$ that are not $n$ or $\mathcal{A}$ 
(the ordering of variables in the determinant is irrelevant for our purposes). The only terms that survive in (\ref{detM}) are those for which the index $A$ takes the row $A = \mathcal{A}$. This is so because when $B = \mathcal{A}$ the terms cancels due to (\ref{dAdA}), and when the row $\mathcal{A}$ does not fall in $A,B$ we can use (\ref{dpsidA}) in the factors
\begin{equation}
\left( \frac{ \delta^2 \mathcal{L} }{ \delta \hat{\Psi}_{\hat{A}} \delta n } \right)
\left( \frac{ \delta^2 \mathcal{L} }{ \delta \hat{\Psi}_{\hat{B}} \delta \mathcal{A} } \right) =
\left( \frac{ \delta^2 \mathcal{L} }{ \delta \hat{\Psi}_{\hat{A}} \delta n } \right)
\left( \frac{ \delta^2 \mathcal{L} }{ \delta \hat{\Psi}_{\hat{B}} \delta n } \right) \,,
\end{equation}
but this is symmetric in $\hat{A},\hat{B}$. Hence, we have
\begin{eqnarray}
\det M &=&
\epsilon_{\mathcal{A} \hat{B} \hat{C} \cdots \hat{D}}
\left( \frac{ \delta^2 \mathcal{L} }{ \delta \mathcal{A} \delta n } \right)
\left( \frac{ \delta^2 \mathcal{L} }{ \delta \hat{\Psi}_{\hat{B}} \delta \mathcal{A} } \right)
W_{\hat{C} \cdots \hat{D}} 
\nonumber
\\ &=&
\left( \frac{ \delta^2 \mathcal{L} }{ \delta n \delta n } \right)
\epsilon_{\hat{B} \hat{C} \cdots \hat{D}}
\left( \frac{ \delta^2 \mathcal{L} }{ \delta \hat{\Psi}_{\hat{B}} \delta n } \right)
W_{\hat{C} \cdots \hat{D}}
\nonumber
\\ &=&
\left( \frac{ \delta^2 \mathcal{L} }{ \delta n \delta n } \right)
\mbox{Minor}(M)_{\mathcal{A}\mathcal{A}} \,.
\label{detminor}
\end{eqnarray}
In the second line we have used (\ref{dpsidA}), and in the third line we have identified the minor of $M$ in the $\mathcal{A}\mathcal{A}$ position. Thus, the determinant of $M$ is proportional to its $\mathcal{A}\mathcal{A}$ minor. Due to this, the entry $[M^{-1}]_{\mathcal{A}\mathcal{A}}$ lacks any contribution coming from the kinetic terms of the Lagrangian, which are the only terms that bring dependence on $\omega$ (indeed, $[M^{-1}]_{\mathcal{A}\mathcal{A}}$ lacks the contribution of most terms of the Lagrangian). The only remainder in this entry is the coefficient, which, in Fourier transform, is
\begin{equation}
\frac{ \delta^2 \mathcal{L} }{ \delta n \delta n } =
\alpha_{67} k^4 \,.
\label{d2L}
\end{equation}
Therefore, we have the propagator
\begin{equation}
\langle \mathcal{A}\mathcal{A} \rangle = 
\frac{1}{\alpha_{67} k^4} \,,
\end{equation}
as shown in (\ref{propagators}).

The other propagator involving the $\mathcal{A}$ field is $\langle\mathcal{A} n\rangle$. We compute $\text{Minor}(M)_{\mathcal{A}n}$ by eliminating row $\mathcal{A}$ and column $n$ from the determinant (\ref{detM}),
\begin{eqnarray}
\mbox{Minor}(M)_{\mathcal{A}n} &=&
\epsilon_{\hat{B} \hat{C} \cdots \hat{D}}
\left( \frac{ \delta^2 \mathcal{L} }{ \delta \hat{\Psi}_{\hat{B}} \delta \mathcal{A} } \right)
W_{\hat{C} \cdots \hat{D}} 
\nonumber
\\ 
&=&
\epsilon_{\hat{B} \hat{C} \cdots \hat{D}}
\left( \frac{ \delta^2 \mathcal{L} }{ \delta \hat{\Psi}_{\hat{B}} \delta n } \right)
W_{\hat{C} \cdots \hat{D}} 
= \mbox{Minor}(M)_{\mathcal{A}\mathcal{A}} \,, 
\label{minorcross}
\end{eqnarray}
hence
\begin{equation}
\langle n \mathcal{A} \rangle = \langle \mathcal{A}\mathcal{A} \rangle = 
\frac{1}{\alpha_{67} k^4} \,.
\label{finirregular}
\end{equation}
Any other propagator involving $\mathcal{A}$ is zero. Let $\phi$ denote any field that is not $\mathcal{A}$ or $n$. The minor of $\mathcal{A}\phi$ is computed by eliminating row $\mathcal{A}$ and column $\phi$,
\begin{eqnarray}
\mbox{Minor}(M)_{\mathcal{A}\phi} &=&
\epsilon_{\hat{A} \hat{B} \cdots \hat{D}}
\left( \frac{ \delta^2 \mathcal{L} }{ \delta \hat{\Psi}_{\hat{A}} \delta n } \right)
\left( \frac{ \delta^2 \mathcal{L} }{ \delta \hat{\Psi}_{\hat{B}} \delta \mathcal{A} } \right)
W^*_{\cdots \hat{D}}
\nonumber
\\  
&=&
\epsilon_{\hat{A} \hat{B} \cdots \hat{D}}
\left( \frac{ \delta^2 \mathcal{L} }{ \delta \hat{\Psi}_{\hat{A}} \delta n } \right)
\left( \frac{ \delta^2 \mathcal{L} }{ \delta \hat{\Psi}_{\hat{B}} \delta n } \right)
W^*_{\cdots \hat{D}} \,,
\label{minorother}
\end{eqnarray}
where $W^*_{\cdots \hat{D}}$ is $W_{\hat{C}\cdots \hat{D}}$ with the factor $\displaystyle \frac{ \delta^2 \mathcal{L} }{ \delta \hat{\Psi}_{\hat{C}} \delta \phi }$ eliminated. The derivatives in the last line are symmetric in $\hat{A}\hat{B}$. Hence we have $\langle \mathcal{A} \phi \rangle = 0$.

A natural question is whether a different gauge-fixing condition could change the situation for the propagators of the $\mathcal{A}$ field. In order for this to happen, the structure (\ref{nAterms}) should be modified. Remarkably, by introducing a functional dependence of the gauge-fixing condition $\chi^i$ on $n$, which is allowed by the BFV formalism, the structure (\ref{nAterms}) is altered since the term $\pi_i \chi^i$ in (\ref{genhamiltonian}) generates additional quadratic terms that depend on $n$. There are other places where dependence on $n$ is generated by the gauge fixing, but they are of higher order in perturbations and hence do not contribute to the propagators. It turns out that the new terms coming from $\pi_i \chi^i$ do not modify the propagators $\langle\mathcal{A}\mathcal{A} \rangle$ and $\langle \mathcal{A} n\rangle$, which can be checked by direct computations. This is the only possibility for the choice of the gauge-fixing condition that could alter the irregular propagators, but at the end it does not happen.

The other sector where irregular propagators arise is on the ghosts $\eta,\bar{\eta}$, which are also associated with the measure of the second-class constraints. In this case the result is more obvious, since the sector $\eta,\bar{\eta}$ forms an isolated block of the $M$ matrix. The only entry where $\eta$ and $\bar{\eta}$ arise is in the term $\displaystyle \bar{\eta} \{ \theta_1 \,, \theta_2 \} \eta $ of (\ref{pathintegralguagefix}), such that the $\langle \bar{\eta} \eta \rangle$ propagator is directly the inverse of this entry of $M$.  In turn, in the linearized theory, this factor becomes
\begin{equation}
\{ \theta_1 \,, \theta_2 \} =
  \frac{ \delta \theta_1 }{ \delta n } =
  \frac{ \delta \mathcal{H}_0^{(2)} }{ \delta n \delta n } =
  \frac{ \delta \mathcal{L}^{(2)} }{ \delta n \delta n } \,.
\label{irregulardet}
\end{equation}
In the last equality we have used (\ref{nAterms}). Thus, this factor is the same as (\ref{d2L}), and the propagator for the ghosts is
\begin{equation}
\langle \eta \bar{\eta} \rangle = \frac{1}{ \alpha_{67} k^{4}} \,.
\end{equation}
Therefore, we have again an irregular propagator that is a consequence of the structure of the second-class constraints.

We stress the origin of the terms that depend on $\eta,\bar{\eta}$, which is the factor $\sqrt{\det\{ \theta_p \,, \theta_q \}}$ of the measure of the second-class constraints. Although we have been able to incorporate this factor into the Lagrangian, its form is robust in two senses: first, the gauge-fixing condition represented by the choice of $\chi^i$ plays no role in this factor; hence, one cannot modify it by a different gauge. Second, there cannot be time derivatives in this factor since the measure is defined strictly in terms of constraints, which by definition do not depend on time derivatives. Moreover, the ghosts $\eta,\bar{\eta}$ are not canonical variables (unlike the BFV ghosts); hence, there is no place for time derivatives of them coming directly from kinetic terms.\footnote{All these features apply also to the terms involving the $\mathcal{A}$ field, but since it couples to canonical fields in the linearized theory, the rest of the Lagrangian is relevant to determine its propagators.}

The presence of the field variables $\mathcal{A}$ and $ \eta, \bar{\eta}$ is a consequence of the formalism of quantization, which requires the measure. In appendix B we comment on an alternative quantization based on the solution of the second-class constraints. The main obstacle with this approach is that although under certain considerations a local Lagrangian can be given, the canonical Lagrangian is essentially nonlocal.


\paragraph{Absence of nested irregular loops}
Our concern here are the loops whose all internal lines are irregular propagators, which we call irregular loops for short. We want to show that nested irregular loops are not possible. To clarify, we ilustrate the idea of nested loops in figure 1.
\begin{figure}[H]
 \begin{center}
 \includegraphics[scale=.3]{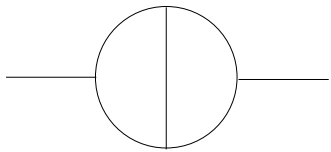}
 \end{center}
\caption{\small An example of nested loops. The integrals over the two loops are not independent}
\end{figure}

\noindent Since the Lagrangian (\ref{pathintegralguagefix}) is linear in $\mathcal{A}$ and bilinear in $\eta\bar{\eta}$, the only possibility is to have vertices with a single $\mathcal{A}$ leg and vertices with a leg for $\eta$ and a leg for $\bar{\eta}$, independently of the number of legs they have for the other fields. Given a diagram or subdiagram, let $V_{\mathcal{A}}$ denote the number of vertices having the $\mathcal{A}$ leg, and $V_{\eta\bar{\eta}}$ the same for the $\eta\bar{\eta}$ ghosts. Since these are the only vertices involving the $\mathcal{A},\eta,\bar{\eta}$ fields, these numbers are related to the corresponding lines according to
\begin{equation}
\begin{split}
 &
 V_{\mathcal{A}} = 
 2 I_{\mathcal{A}\mathcal{A}} + I_{\mathcal{A}n} + E_{\mathcal{A}} \,,
 \\ &
 V_{\eta\bar{\eta}} = I_{\eta\bar{\eta}} + 2 E_{\eta} \,,
\end{split}
 \label{vertices}
\end{equation}
where $E_{\mathcal{A}}$ is the number of external $\mathcal{A}$ legs and $E_{\eta}$ the number of external $\eta$ legs (which must be the same of the external $\bar{\eta}$ legs). We have left open the possibility of having external legs for the ghosts with the aim of applying the analysis for subdiagrams. In the general case, we have a diagram or subdiagram with $L$ loops, $I_R + I_Q$ internal lines, where $I_R$ are the regular propagators and $I_Q$ the irregular ones, $V_{\mathcal{A}}$ and $V_{\eta\bar{\eta}}$ vertices, and the rest are the $V$ vertices. The topological identity is
\begin{equation}
 L - 1 = I_R + I_Q - ( V_{\mathcal{A}} + V_{\eta\bar{\eta}} + V ) \,.
\label{loops}
\end{equation}
Now we consider a diagram whose all internal lines are irregular propagators. In this case we have $I_R = V = 0$, since it has no regular propagators and the vertices $V$ are not allowed since they have no legs to join with the irregular propagators. The lines corresponding to irregular propagators are $I_Q = I_{\mathcal{A}\mathcal{A}} + I_{\mathcal{A}n} + I_{\eta\bar{\eta}}$. By substituting this and (\ref{vertices}) in (\ref{loops}), we obtain
\begin{equation}
 L - 1 =  
 - ( I_{\mathcal{A}\mathcal{A}} + E_{\mathcal{A}} + 2 E_{\eta} ) \,.
\end{equation}
But this is only possible, besides the tree level, for $L=1$ and $I_{\mathcal{A}\mathcal{A}} = E_{\mathcal{A}} = E_{\eta} = 0$. Therefore, diagrams or subdiagrams with all internal lines given by irregular propagators can be formed only at one loop, only with the propagators $\langle \mathcal{A}n \rangle$ or $\langle \eta\bar{\eta} \rangle$ and necessarily with external legs corresponding to regular lines. A corollary is that nested irregular loops can not be formed since they would be, at least, of two loops. This is what we wanted to prove.

Multiple irregular loops can arise in the form of several irregular one-loop subdiagrams, but these subdiagrams must be connected by regular propagators ($E_{\mathcal{A}} = E_{\eta} = 0$). Hence, they cannot be nested. We illustrate this kind of multiple irregular one-loop subdiagram in figure 2.
\begin{figure}[H]
 \begin{center}
 \includegraphics[scale=.4]{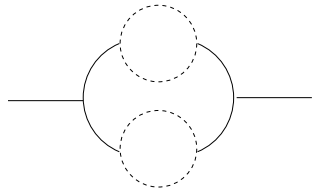}
 \end{center}
 \caption{\small A diagram with multiple irregular one-loops. The irregular propagators are represented generically by dotted lines.}
\end{figure}
\noindent Since the irregular loops cannot be nested, they factorize in the computations of diagrams that is, their respective integrals are independent and multiply each other. Another possibility is that irregular loops can be nested with other loops with some irregular lines, but not all of their lines. We illustrate an example where the $\langle \mathcal{A} \mathcal{A} \rangle$ propagator is used as additional irregular line in figure 3.
\begin{figure}[H]
 \begin{center}
 \includegraphics[scale=.31]{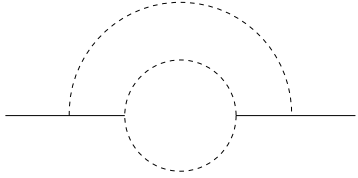}
 \end{center}
 \caption{\small Nested loops with some irregular lines. Only the inner loop is an irregular loop.}
\end{figure}
\noindent In this case the regular propagators arising in the outer loop render the integration in $\omega$ finite for that loop (we discuss this below). Therefore, the degree of divergence on $\omega$ of this diagram is the same of the inner irregular loop.

An additional remarkable feature of the irregular one-loop diagrams is that they can be formed with multiple $\langle \mathcal{A} n \rangle$ or $\langle \eta\bar{\eta} \rangle$ propagators, but not with mixing between them. This is just because there is no propagator connecting $\mathcal{A}$,$n$ legs with $\eta,\bar{\eta}$ legs. Hence there is no way to close a loop when one attempts to join a vertex that has the $\mathcal{A},n$ legs with a vertex that has the $\eta,\bar{\eta}$ legs.

\paragraph{Divergence of the irregular one-loops}
The irregular propagators have all the same form; hence, we can give a general evaluation of irregular one-loops formed with them. These one-loops  lead to the integral
\begin{equation}
  \left[ \cdots \right] 
  \int d\omega \int d^2k E(k^i; p^i,\ldots) \,,
\end{equation}
where $E(k^i; p^i,\ldots)$ is some expression of the spatial momentum $k^i$ circulating along the loop and the external momenta $p^i,\ldots$. The ellipsis at the left is the rest of the diagram that does not enter in the integration over the loop. In the integrand there is no dependence on $\omega$ because the irregular loops do not depend on it and the vertices in this formalism have no time derivatives. With respect to the integration on $k^i$, the propagators $\langle \mathcal{A} n \rangle$ and $\langle \eta \bar{\eta} \rangle$ given in (\ref{propagators}) behave as regular propagators (also $\langle \mathcal{A} \mathcal{A} \rangle$). The scalings on $k^i$ of these propagators and the vertices may lead to a cancellable divergence, if any. This is in agreement with the power-counting renormalizability. The integration on $\omega$ leads to a linear divergence that multiplies the whole loop and hence multiplies the whole diagram. Thus, the irregular loops are potentially dangerous due to this multiplicative divergence on $\omega$.

The irregular one-loops are the only loops that exhibit this behavior in their integration on $\omega$. Indeed, it is enough that a loop has at least one regular propagator to avoid any divergence coming directly from the $\omega$ direction. We may check this starting with the worst case, which is the case of loops that have a single regular propagator of the lower negative scaling in $\omega$, $\omega^{-1}$. These propagators are, for example, $\langle h_{ij} p^{kl} \rangle$ and $\langle p^{ij} n \rangle$. They involve a single canonical momentum. Since the only dependence the integrand has in $\omega$ comes from this propagator, one might face a logarithmic divergence in $\omega$. But, these propagators are odd in $\omega$. Hence, loops with a single propagator of this kind are automatically zero due to the integration in the entire domain of $\omega$. The next level are the loops that have a single regular propagator with scaling $\omega^{-2}$, or two of the previous propagators forming an even integrand, yielding also a total scaling of $\omega^{-2}$. The integration in $\omega$ does not lead to divergences in these cases.


\paragraph{Exact cancellations between irregular one-loops}
We have shown that irregular loops can be formed only as isolated one-loops, and only with $\langle \mathcal{A}n \rangle$ or $\langle \eta \bar{\eta} \rangle$ propagators, but not with mixing between them. Now we show that these features lead to a general exact cancellation between the two classes of irregular one-loops.\footnote{We are very grateful to Sergey Sibiryakov and Mario Herrero-Valea, whose observations motivated us to study the cancellations thoroughly.}

To form the irregular loops with the propagator $\langle \mathcal{A} n \rangle$ we require that all vertices in the loop have an $\mathcal{A}$ leg and at least one $n$ leg, such that a pair of $\mathcal{A},n$ legs combine with the irregular propagators. These vertices come from the term $\mathcal{A} \theta_1$ in (\ref{pathintegralguagefix}). Hence, the terms of $\theta_1$ corresponding to these vertices must have, at least, one factor of $n$. At arbitrary order in perturbations, the general form of such terms is
\begin{equation}
\gamma L_1 n L_2 n \cdots R \,, 
\label{ntheta}
\end{equation}
where $L_1,L_2,\ldots$ are spatial differential operators, $R$ stands for the rest of field factors that do not depend on $n$, and $\gamma$ represents a generic coupling constant. Thus, vertices entering in the irregular $\mathcal{A} n$ one-loops come from terms of the form
\begin{equation}
\gamma \mathcal{A} L_1 n L_2 n \cdots R \,.
\label{AnR}
\end{equation}

The irregular $\eta \bar{\eta}$ loops are formed with vertices that come from the factor
\begin{equation}
\bar{\eta} \{ \theta_1 \,, \theta_2 \} \eta =
\bar{\eta} \frac{\delta \theta_1}{\delta n} \eta \,.
\label{measureeta}
\end{equation}
Since this is generated through the derivative $\displaystyle \frac{\delta \theta_1}{\delta n} $, the set of terms of $\theta_1$ that contribute to (\ref{measureeta}) is just the set of terms represented in (\ref{ntheta}). Given a term of the form (\ref{ntheta}), we obtain the terms it generates for $\eta,\bar{\eta}$,
\begin{equation}
\gamma \bar{\eta} \left( 
L_1 \eta L_2 n \cdots + L_1 n L_2 \eta \cdots + \cdots + L_1 n L_2n \cdots \eta  
\right) R \,.
\label{etabareta}
\end{equation} 
Now the matching between irregular one-loops comes out. On the side of the $\mathcal{A} n$ loops, for each vertex (\ref{AnR}) one must choice one of the $n$ legs to contract with the propagators in the loop. On the side of the $\eta\bar{\eta}$ loops, each term in (\ref{etabareta}) has only one possibility of contracting the $\eta,\bar{\eta}$ legs, but the different terms in (\ref{etabareta}) are equivalent to the different choices of $n$ in (\ref{AnR}) because it comes from its derivative with respect to $n$. Therefore, for each vertex (\ref{AnR}) that has been used with a choice of one $n$ leg in the loop, there is an equivalent vertex in (\ref{etabareta}) whose $\eta,\bar{\eta}$ legs have entered in the loop. Notice that the rest of the factors are exactly the same for both cases, since the only difference is the exchanging of the pairs $\mathcal{A},n$ and $\eta,\bar{\eta}$ between vertices. Moreover, the two irregular propagators are identical, $\langle \mathcal{A} n \rangle = \langle \eta \bar{\eta} \rangle$. Thus, each irregular one-loop formed with the $\langle \mathcal{A} n \rangle $ propagator has a counterloop formed with the $\langle \eta \bar{\eta} \rangle$ propagator that is numerically identical. The only difference is the relative minus sign due to $\eta,\bar{\eta}$ are fermions. Therefore, the two irregular one-loops cancel themselves exactly; the multiplicative divergences in $\omega$ disappear.

We present two examples of one-loop diagrams showing the exact cancellation using three-leg vertices. In appendix A we show the expansion of the Lagrangian at cubic order in all variables. Here, it is enough to take only the cubic terms that are proportional to the same coupling constant, since the coupling constants are independent and cancellations occur separately for them. The first example is given by diagrams with two external $n$ legs and one irregular loop. We take the vertices proportional to the coupling constant $\alpha_4$. To form the cubic terms that contribute to the diagrams, we require the terms of $\theta_1$ that are quadratic in $n$ and the terms of $\{\theta_{1},\theta_{2}\}$ that are linear in $n$. These terms are
\begin{eqnarray}
 \theta_{1} &\sim&
   2 \alpha_4 \partial_{ij} n \partial_{ij} n
 - 2 \alpha_4 \Delta n\Delta n  \,,
 \\ 
 \{\theta_{1},\theta_{2}\} &\sim&
   4 \alpha_4 \partial_{ij} n \partial_{ij}
 - 4 \alpha_4 \Delta n\Delta  \,.
 \qquad
\end{eqnarray} 
Terms in the last line arise with a doubled relative weight because they are the derivatives of the terms of the first line, as we discussed above. The vertices in Fourier space are
\begin{eqnarray}
 \begin{minipage}{5em}
 \includegraphics[scale=.2]{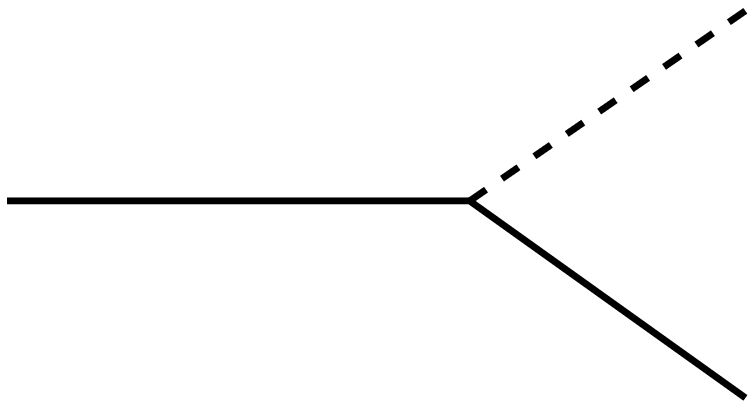}
 \end{minipage}
  &=& 
  2\alpha_{4}\left( (p_{i} k_{i})^2 - p^2k^2\right)  \,,
  \label{Ann}
 \\[1ex]
 \begin{minipage}{5em}
 \includegraphics[scale=.2]{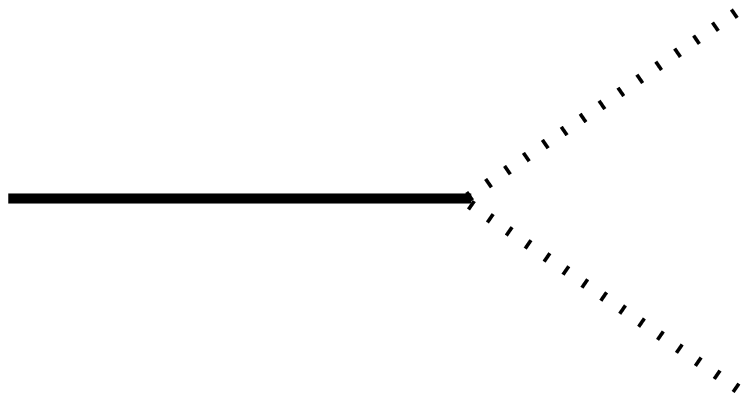}
 \end{minipage}
  &=&
  4 \alpha_{4} \left( (p_{i} k_{i})^2 - p^2k^2\right) \,.
  \label{netaeta}
\end{eqnarray}
In vertex (\ref{Ann}) solid lines stand for the $n$ legs with spatial momentum $p_i$ and $k_i$, and the dashed line is for the $\mathcal{A}$ leg. In vertex (\ref{netaeta}) the solid line is for $n$ with momentum $p_i$, the dashed lines are for the ghosts $\eta,\bar{\eta}$, and the line of $\eta$ carries momentum $k_i$. Diagrams with one irregular loop and two external $n$ legs require the combination of two of the above vertices. The two vertices (\ref{Ann}) yield four possible combinations according to which leg of $n$ is used as internal or external, and the four diagrams give the same result. The two vertices (\ref{netaeta}) allow for a single combination. Thus, we have that numerical factors compensate in the sum of all diagrams,
\begin{equation}
  ( 4 \times ) \,
  \begin{minipage}{7em}
  \includegraphics[scale=.2]{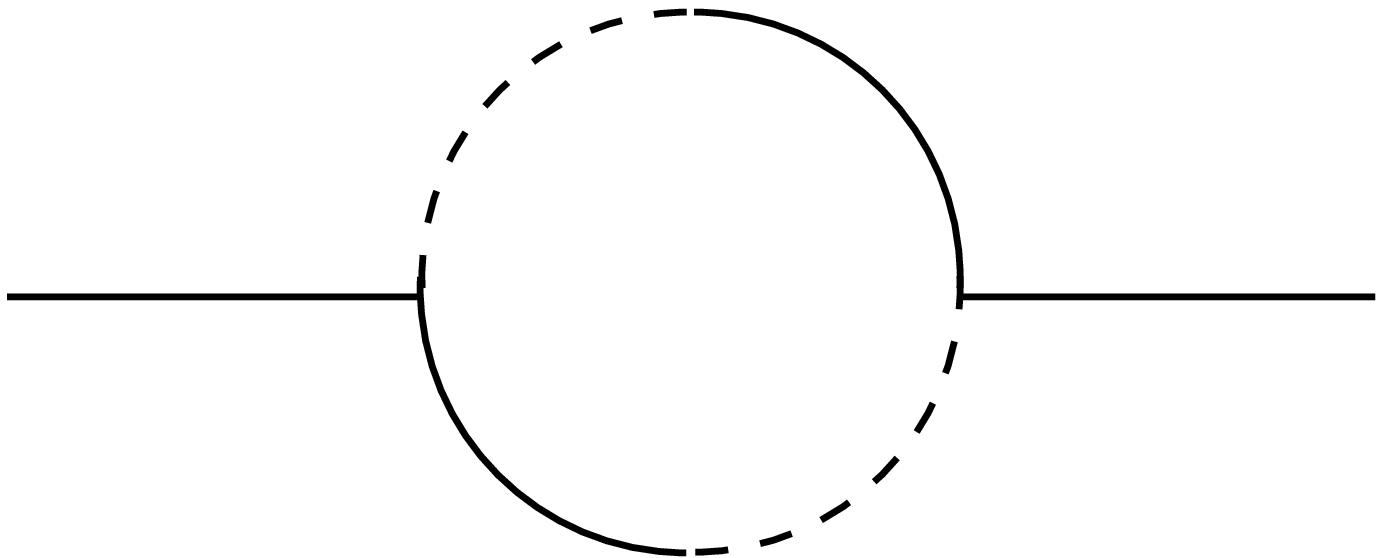}
  \end{minipage}
  \quad + \quad 
  \begin{minipage}{7em}
  \includegraphics[scale=.2]{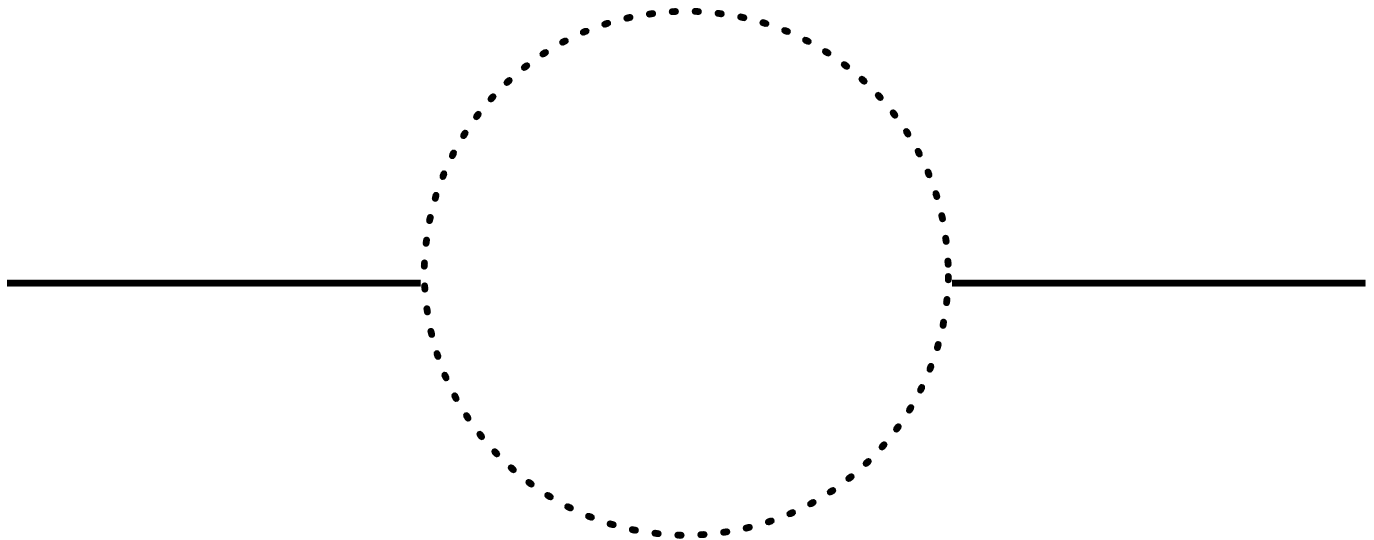} 
  \end{minipage}
  \quad =\, 0 \,.
\end{equation}

The second example is given by the analogous diagrams with two external $h^T$ legs. To form such diagrams we require terms in $\theta_1$ that are linear in $h^T$ and $n$. In this case we take the terms of the Lagrangian proportional to $\alpha_3$ (see appendix A). The vertices are
\begin{eqnarray}
 \begin{minipage}{5em}
 \includegraphics[scale=.2]{./vertex.eps}
 \end{minipage}
  &=& 
  2\alpha_{3}p^{2}\left(k^2 - k_{i}p_{i}\right) \,,
  \label{hnA}
  \\[1ex]
  \begin{minipage}{5em}
  \includegraphics[scale=.2]{./vertex2.eps}
  \end{minipage}
  &=&
  2 \alpha_{3} p^{2}\left(k^{2} - k_{i}p_{i}\right) \,.
  \label{hetaeta}
\end{eqnarray}
In (\ref{hnA}) solid lines are for $h^T$ (momentum $p_i$) and $n$ (momentum $k_i$), and the dashed line is for $\mathcal{A}$. In the vertex (\ref{hetaeta}) the solid line is for $h^T$ with momentum $p_i$, whereas $\eta$ has momentum $k_i$. The relative numerical weight between both vertices is the same since they come from terms of $\theta_1$ that are linear in $n$. Both kinds of vertices allow for a unique combination yielding the diagrams of interest. Therefore, we have that
\begin{equation}
  \begin{minipage}{7em}
  \includegraphics[scale=.2]{./loopdivergence.eps}
  \end{minipage}
  \quad + \quad
  \begin{minipage}{7em}
  \includegraphics[scale=.2]{./loopdivergenceghost.eps}
  \end{minipage}
  \quad = \, 0\,.
\end{equation}
These two examples evidence the exact matching between the possibilities of using vertices with $\mathcal{A},n$ legs and vertices with $\eta,\bar{\eta}$ legs. In the first example an $nn\mathcal{A}$ vertex has two possibilities of combination. This is compensated with the extra factor of $2$ that has the $n\eta\bar{\eta}$ vertex, since it is its derivative, and admits a unique combination. In the second example both vertices have the same relative weight and admit a unique combination.

We emphasize that the sectors $\mathcal{A} \theta_1$ and $\bar{\eta} \{ \theta_1 \,, \theta_2 \} \eta$ of the Lagrangian in (\ref{pathintegralguagefix}) are in general different and they have very involved expressions in perturbations. The matching between them occurs only for the vertices involved in irregular loops, but for other diagrams they give different contributions in general.


\section{3+1 dimensions}
The relationship between the measure of the second-class constraints and the irregular propagators can be translated to the case of a foliation of $3+1$ dimensions. Thanks to the previous analysis, we may discuss this on very general grounds, without necessity of entering in detailed computations (we presented the BFV quantization of the $3+1$ nonprojectable theory in \cite{Bellorin:2021udn}).

In $3+1$ dimensions the Lagrangian and the classical primary Hamiltonian are of $z=3$ order; that is, they have terms up to sixth order in spatial derivatives. Constraint $\theta_1$ is still a derivative of $\mathcal{H}_0$, getting the $z=3$ order. The gauge-fixing condition must be adapted to the dimension and the anisotropic scaling, but it remains as a functional of $N_i$, $\pi_i$, and $h_{ij}$. Once the measure of the second-class constraints has been incorporated, the dependence of the quantum canonical Lagrangian on $n$ and $\mathcal{A}$ keeps the structure (\ref{nAterms}). Therefore, all the analysis leading to (\ref{detminor}), (\ref{minorcross}), and (\ref{minorother}) holds in the same way. The field $\mathcal{A}$ acquires irregular propagators, with a scaling $k^{-6}$ on spatial momentum. In the case of the ghosts $\eta,\bar{\eta}$, the result (\ref{irregulardet}) is also kept. It is again a consequence of the structure of the second-class constraints. Thus, the same conclusion is valid for these ghosts; they acquire irregular propagators. These facts lead to the result that the propagators of the fields $\mathcal{A}$ and $\eta,\bar{\eta}$ form irregular one-loops that have the multiplicative subdivergence in $\omega$. Since the functional relationship between the $\mathcal{A} \theta_1$ and $\bar{\eta} \{ \theta_1 \,, \theta_2 \} \eta$ terms is kept in $3+1$ dimensions, it is clear that the exact cancellations between the irregular one-loops hold in this case.


\section{Conclusions}
We have performed a study of loops and divergences on the nonprojectable Ho\v{r}ava theory based on the Feynman rules derived from the BFV quantization. This is a Hamiltonian formalism that allows us to pose a local canonical Lagrangian with a suitable gauge-fixing condition and the second-class constraints. We have shown that there are irregular one-loops producing multiplicative subdivergences in the direction of the frequency. These loops are formed with irregular propagators of fields associated with the measure of the second-class constraints. We emphasize that the local gauge-fixing condition yields regular propagators for the canonical fields, but not for these fields. The second-class constraints have a structure that leads unavoidably to the irregular propagators. We have discussed that changing the gauge-fixing condition has no influence on them.

Although the multiplicative subdivergence of each irregular one-loop is potentially dangerous, we have found the central result that irregular loops cancel exactly between them, thanks to a perfect matching between the fields and ghosts producing the irregular loops. To arrive at this cancellation, we have found some particular features of the irregular loops. Remarkably, only isolated irregular one-loops can be formed. They cannot be nested. There are only two kinds of irregular loops, one for fields and the other one for ghosts. They cancel themselves exactly due to a functional relationship between them: the number of ghost loops that can be formed is equivalent to the combinatorics of the loops made from fields.

We have discussed that only the irregular loops produce the multiplicative subdivergence. Hence, their cancellation is an essential step towards the renormalization of the theory. Irregular propagators still contribute to diagrams in other ways (combined with regular propagators), but the divergences they produce are qualitatively the same of the loops with regular propagators. Therefore, our analysis show that the behavior of divergences is equivalent to the projectable theory \cite{Barvinsky:2015kil}. Counterterms are local and their maximal order is the same as the one of the bare Lagrangian. 

Further analysis is required to understand the influence of the gauge symmetry on the counterterms. This would be the final study to arrive at the renormalization of the theory. It is well known that for gauge theories the BRST symmetry is the adequate framework to characterize the counterterms and to establish the complete renormalizability. In the case of the nonprojectable Ho\v{r}ava theory, a thorough analysis on the relationship between the BRST symmetry, the second-class constraints, and locality is required. We leave this for future work.

\section*{Acknowledgments}
The authors wish to thank Sergey Sibiryakov and Mario Herrero-Valea for very useful discussions. C.B.~is partially supported by Grant No. CONICYT PFCHA/DOCTORADO BECAS CHILE /2019 -- 21190960. B.~D.~is partially supported by the Universidad de Antofagasta Grant No. PROYECTO ANT2156, Chile. C.B.~is a graduate student in the ``Doctorado en F\'isica Menci\'on F\'isica-Matem\'atica" Ph.D. program at the Universidad de Antofagasta.

\begin{appendices}
\section{Cubic order}
For the sake of simplicity, here we show only the $z=2$ terms:
\begin{eqnarray}
&&
\mathcal{H}_{0}^{(3)} =
2p^{jk} h_{ki} p^{ij}
+ 2 \bar{\sigma} pp^{ij}h_{ij}
+ \left(n-\frac{1}{2} h \right)\left(p^{ij} p^{ij} 
+\bar{\sigma} p^2\right)
+\alpha_{1}\left[\left(n+\frac{1}{2} h \right)(R^{(1)})^{2} 
\right.
\nonumber 
\\  
&& \left.   
-2R^{(1)} h_{kl} R^{(1)}_{kl}-2R^{(1)}R^{(2)}\right]
+\alpha_{3}\left(\partial_{ij}h_{ij}\partial_{l}n\partial_{l}n
-\Delta h\partial_{l}n\partial_{l}n\right)
+\alpha_{4}\partial_{i}n\partial_{i}n\Delta n
\nonumber
\\
&&
+\alpha_{5}\left[
R^{(1)}\left(\frac{h}{2}\Delta n-\partial_{k}n\partial_{k}n
-h_{kl}\partial_{kl} n
+ \chi^{l}_{kk}\partial_{l}n\right)
+\Delta nR^{(2)}\right.\nonumber\\
&&
-\left.\frac{1}{2}\Delta nh_{ij}\left(\partial_{ki} h_{kj}
-\partial_{ij} h
-\Delta h_{ij}+\partial_{kj}h_{ki}
\right)\right]
- \alpha_{6}\left( n\partial_{ij}n\partial_{ij}n
+ 2\partial_{ij}n \partial_{i}n\partial_{j}n
\right.
\nonumber\\
&& \left.  
- \frac{h}{2}\partial_{ij} n\partial_{ij}n
+ 2\partial_{ij}n\partial_{m}n\chi^{m}_{ij}
+ 2h_{jl}\partial_{il}n\partial_{ij}n
\right)
- \alpha_{7}\left( n\Delta n\Delta n - \frac{h}{2}\Delta n\Delta n
\right.
\nonumber\\
&&    
+ 2\Delta n\partial_{j}n\partial_{j}n
+ 2\Delta n\partial_{k}n\delta^{jl}\chi^{k}_{jl}
+ 2h_{jl}\Delta n\partial_{jl} n \Big) \,,
\end{eqnarray}
where
\begin{eqnarray}
\chi^{i}_{kl} &=&
\partial_{k}h_{li} - \frac{1}{2}\partial_{i}h_{kl} \,,
\\
R^{(1)}_{ij} &=&
\frac{1}{2}\partial_{i}\partial_{k}h_{kj}
+ \frac{1}{2}\partial_{j}\partial_{k}h_{ki}
- \frac{1}{2}\partial_{i}\partial_{j}h
- \frac{1}{2}\Delta h_{ij}
\,,
\\
R^{(2)}&=&
- \partial_{k}h_{kl}\partial_{i}h_{li}
+ \partial_{k}h_{kl}\partial_{l}h
- 2h_{kl}\partial_{k}\partial_{i}h_{li}
+ h_{ij}\partial_{i}\partial_{j}h
\nonumber\\&&
+\frac{3}{4}\partial_{i}h_{kl}\partial_{i}h_{kl}
+h_{kl}\Delta h_{kl}
-\frac{1}{4}\partial_{i}h\partial_{i}h
-\frac{1}{2}\partial_{j}h_{ik}\partial_{k}h_{ji}
\,.
\end{eqnarray}
Other expansions are
\begin{eqnarray}
&&
\theta_{1}^{(2)} = 
p^{ij}p^{ij} + \bar{\sigma} p^2 +(-\alpha_{6}+2\alpha_{4}-4\alpha_{7})\partial_{ij}n\partial_{ij}n 
+(\alpha_{7}-2\alpha_{4}-2\alpha_{6})\Delta n\Delta n
\nonumber\\ &&
+ \alpha_{1} \Big( \partial_{ij}h_{ij}\partial_{kl}h_{kl}
-2 \Delta h\partial_{ij} h_{ij}
+ \Delta h\Delta h
\Big)
+ (2\alpha_{3}-2\alpha_{5}) \Big( \Delta n\Delta h 
- \Delta n\partial_{ij}h_{ij}
\nonumber\\ &&
- \partial_{k}n\partial_{kij} h_{ij} \Big)
+(2\alpha_{3}-2\alpha_{5}+\alpha_{67})\partial_{k}n\Delta\partial_{k}h
+(-2\alpha_{6}-4\alpha_{7})\partial_{i}h_{ij}\partial_{j}\Delta n
\nonumber\\ &&
-2\alpha_{6} \Big(\Delta h_{kl}\partial_{kl} n
+ 2 \partial_{l}h_{ij}\partial_{lij} n \Big)
+ \alpha_{67} \Big( 
  h\Delta^{2}n
+ 2 \partial_{k}h\Delta\partial_{k}n
+ 2 \partial_{ij}n\partial_{ij}h
-4 \partial_{jk} h_{ik} \partial_{ji} n
\nonumber\\ &&
-2 \partial_{i}n\Delta\partial_{k}h_{ik}
-4 \Delta\partial_{k}n\partial_{k}n
-4 h_{ij}\Delta\partial_{ij} n \Big)
+ \alpha_5 \Big( 
 \frac{1}{2} h\Delta\partial_{ij} h_{ij}
+\frac{1}{2}\partial_{k}h\partial_{kij} h_{ij}
\nonumber \\ &&
+2h_{ij}\partial_{ij} \Delta h
+\Delta\partial_{k}h_{kl}\partial_{l}h
+2\partial_{jk} h_{kl} \partial_{jl} h
+2\partial_{k}h_{ij}\partial_{kij}h
-\partial_{i}h_{ij}\partial_{jkl} h_{kl}
-h_{ij}\partial_{ijkl} h_{kl}
\nonumber\\
&&
-4\partial_{j}h_{kl}\partial_{jki} h_{li}
-2h_{kl}\Delta\partial_{ki} h_{li}
+\frac{7}{2}\Delta\partial_{i}h_{kl}\partial_{i}h_{kl}
+\frac{3}{2}\partial_{ji} h_{kl}\partial_{ji} h_{kl}
+\Delta h_{ij}\Delta h_{ij}
+h_{ij}\Delta^{2} h_{ij}
\nonumber\\ &&
-\Delta\partial_j h_{ik}\partial_{k}h_{ji}
-\partial_{lj} h_{ik}\partial_{lk} h_{ji}
-\frac{1}{2}h\Delta^{2}h
-\partial_{k}h\partial_{k}\Delta h
-\frac{1}{2}\partial_{ji} h\partial_{ji} h
-n\Delta^{2}h
+n\Delta\partial_{ij} h_{ij}
\nonumber \\ &&
+\Delta h_{ij}\partial_{ij}h 
-2\Delta\partial_{k}h_{kl}\partial_{i}h_{li}
-2\partial_{jk} h_{kl}\partial_{ji} h_{li}
-2\Delta h_{kl}\partial_{ki} h_{li}
+2\partial_{i}h_{ij}\partial_{j}\Delta h \Big) 
\end{eqnarray}

\begin{eqnarray}
&&
\det\{\theta_{1},\theta_{2}\}^{(1)} =
\nonumber \\ &&  
2(-\alpha_{6}+2\alpha_{4}-4\alpha_{7})\partial_{ij}n\partial_{ij}
+ 2(\alpha_{7}-2\alpha_{4}-2\alpha_{6})\Delta n\Delta 
- 4\alpha_{67}\left(\partial_{k}n\Delta\partial_{k}+\Delta\partial_{k}n\partial_{k}\right)
\nonumber\\
&& + 2(\alpha_{3}-\alpha_{5})\left(\Delta h\Delta
- \partial_{ij}h_{ij}\Delta
+ \Delta\partial_{k}h\partial_{k}
- \partial_{kij}h_{ij}\partial_{k}\right) 
+ \alpha_{67}\left(h\Delta^{2}
+ \Delta\partial_{k}h\partial_{k}
\right)
\nonumber\\
&&
+ 2\alpha_{67}\left(\partial_{k}h\Delta\partial_{k}
+ \partial_{ij}h\partial_{ij}\right)
- \alpha_{5}\left(\Delta^{2}h -\Delta\partial_{ij}h_{ij}\right)
- 2(\alpha_{6}+2\alpha_{7})\partial_{i}h_{ij}\partial_{j}\Delta
\nonumber\\
&& - 4\alpha_{67}\left(\partial_{jk}h_{ik}\partial_{ji}
+ h_{ij}\Delta\partial_{ij}\right)
- 2\alpha_{67}\Delta\partial_{k}h_{ki}\partial_{i}
- 2\alpha_{6}\left(\Delta h_{kl}\partial_{kl}
+ 2\partial_{l}h_{ij}\partial_{lij}\right).
\end{eqnarray}

\section*{Appendix B \, Reduced phase space}
One way to avoid dealing with second-class constraints in the quantum theory is to solve them. It is remarkable that, under certain assumptions, one can give the solution of the $\theta_1$ constraint and still arrive at a local canonical Lagrangian. We show this in the following, by considering the linear theory with the dominant modes at the ultraviolet regime. Keeping only the coupling constants of the $z=2$ terms, the linearized constraint becomes
\begin{eqnarray}
\theta_{1}^{(1)} &=&
- \alpha_{5}\Delta^{2}h^{T}
+ 2 \alpha_{67} \Delta^{2}n=0 \,.
\label{linearhamconst}
\end{eqnarray}
With suitable boundary conditions, we may solve it for $n$,
\begin{equation}
n = \frac{\alpha_{5}}{2\alpha_{67}} h^{T} \,.
\label{nsolution}
\end{equation}
Thus, $n$ can be integrated out in the linearized theory, substituting (\ref{nsolution}) everywhere in the path integral. By definition, the measure of the second-class constraint becomes equal to $1$ evaluated on the solution. The path integral (\ref{pathintegralguagefix}) becomes
\begin{equation}
\begin{split}
Z =& 
\int 
\mathcal{D} h_{ij} \mathcal{D} p^{ij} 
\mathcal{D} n^{i} \mathcal{D}\pi_{i}
\mathcal{D} C^{i} \mathcal{D}\bar{\mathcal{P}}_{i} 
\mathcal{D}\bar{C}_{i} \mathcal{D} \mathcal{P}^i 
\exp\left[ i \int dt d^{2}x \Big(
\bar{\mathcal{P}}_{i}^{T} \dot{C}^{T}_i
- \bar{\mathcal{P}}^{L} \Delta\dot{C}^{L}
+ \mathcal{P}^{T}_i \dot{\bar{C}}_{i}^{T}
\right.
\\ &
- \mathcal{P}^{L} \Delta \dot{\bar{C}}^{L} 
- \bar{\mathcal{P}}_{i}^{T}\mathcal{P}^{T}_i
+ \bar{\mathcal{P}}^{L}\Delta\mathcal{P}^{L}
+ 2\bar{C}_{i}^{T}\Delta^{2}C^{T}_i
- 2 \rho_1 \bar{C}^{L}\Delta^{3}C^{L}
- \sigma \left(p^{T}\right)^{2}
\\ &
+ p^{T}\left(\dot{h}^{T}- 2 \bar{\sigma} p^{L}\right)
- \frac{1}{2}p^{L}_i \Delta\dot{h}_{i}^{L}
+ p^{L}\dot{h}^{L}
+ \frac{1}{2}p^{L}_i \Delta p^{L}_i
- \sigma \left(p^{L}\right)^{2}
+ p^{L}_i \Delta n^{T}_i
- 2p^{L}\Delta n^{L}
\\ &
- \frac{\rho_2}{4}  h^{T}\Delta^{2}h^{T}
+ \pi_{i}^{T}\dot{n}^{T}_i
- \pi^{L}\Delta\dot{n}^{L}
- \pi_{i}^{T}\Delta\pi_{i}^{T}
+ \bar{\kappa}\pi^{L}\Delta^{2}\pi^{L}
+ \pi_{i}^{T}\Delta^{2}h_i^{L}
+ 2\lambda\bar{\kappa}\pi^{L}\Delta^{2}h^{T}
\\ &
\left.	
- \rho_1 \pi^{L}\Delta^{2}h^{L}
\Big)\right]
\,.
\end{split}
\end{equation}
Therefore, under these conditions one obtains a local canonical quantum Lagrangian, which is of quadratic order. Regular propagators can be obtained for all quantum modes in this case. Despite this, the locality of the quantum Lagrangian is broken in two ways. The first one is when one includes interactions, such that higher orders in perturbations must be considered. In this case the solution of the $\theta_1$ constraint is required at second order and higher, leading to a nonlocal solution for $n$. Since this affects the vertices where $n$ arises, these vertices get nonlocal expressions. The second one is when one recovers the terms of lower order in derivatives. This is again an obstruction to get a local solution for $n$, 
leading to a nonlocal Lagrangian.

\end{appendices}




\begin{thebibliography}{99}
	\bibitem{Horava:2009uw} 
	P.~Ho\v{r}ava,
	Quantum Gravity at a Lifshitz Point,
	Phys.\ Rev.\ D {\bf 79} 084008 (2009) 
	[arXiv:0901.3775 [hep-th]].
	
	\bibitem{Blas:2009qj} 
	D.~Blas, O.~Pujolas and S.~Sibiryakov,
	Consistent Extension of Ho\v rava Gravity,
	Phys.\ Rev.\ Lett.\  {\bf 104} 181302 (2010)
	[arXiv:0909.3525 [hep-th]].

	\bibitem{Stelle:1976gc}
	K.~S.~Stelle,
	Renormalization of Higher Derivative Quantum Gravity,
	Phys. Rev. D \textbf{16} 953 (1977).
	
	\bibitem{Kluson:2010nf}
	J.~Kluson,
	Note About Hamiltonian Formalism of Healthy Extended Ho\v{r}ava-Lifshitz Gravity,
	JHEP \textbf{07} 038 (2010)
	[arXiv:1004.3428 [hep-th]].
	
	\bibitem{Donnelly:2011df}
	W.~Donnelly and T.~Jacobson,
	Hamiltonian structure of Ho\v{r}ava gravity,
	Phys. Rev. D \textbf{84} 104019 (2011)
	[arXiv:1106.2131 [hep-th]].
	
	\bibitem{Bellorin:2011ff}
	J.~Bellor\'in and A.~Restuccia,
	Consistency of the Hamiltonian formulation of the lowest-order effective action of the complete Ho\v{r}ava theory,
	Phys. Rev. D \textbf{84} 104037 (2011)
	[arXiv:1106.5766 [hep-th]].
	
	\bibitem{Kobakhidze:2009zr}
	A.~Kobakhidze,
	On the infrared limit of Ho\v{r}ava's gravity with the global Hamiltonian constraint,
	Phys. Rev. D \textbf{82} 064011 (2010)
	[arXiv:0906.5401 [hep-th]].
		
	\bibitem{Barvinsky:2015kil}
	A.~O.~Barvinsky, D.~Blas, M.~Herrero-Valea, S.~M.~Sibiryakov and C.~F.~Steinwachs,
	Renormalization of Ho\v{r}ava gravity,
	Phys. Rev. D \textbf{93} 064022 (2016)
	[arXiv:1512.02250 [hep-th]].
	
    \bibitem{Orlando:2009en}
    D.~Orlando and S.~Reffert,
    On the Renormalizability of Ho\v{r}ava-Lifshitz-type Gravities,
    Class.\ Quant.\ Grav.\  {\bf 26} 155021 (2009)
    [arXiv:0905.0301 [hep-th]].
    
    \bibitem{Contillo:2013fua}
    A.~Contillo, S.~Rechenberger and F.~Saueressig,
    Renormalization group flow of Ho\v{r}ava-Lifshitz gravity at low  energies,
    JHEP {\bf 12} 017 (2013)
    [arXiv:1309.7273 [hep-th]].
    
    \bibitem{D'Odorico:2014iha}
    G.~D'Odorico, F.~Saueressig and M.~Schutten,
    Asymptotic Freedom in Ho\v{r}ava-Lifshitz Gravity,
    Phys.\ Rev.\ Lett.\  {\bf 113} 171101 (2014)
    [arXiv:1406.4366 [gr-qc]].
    
    \bibitem{Benedetti:2013pya}
    D.~Benedetti and F.~Guarnieri,
    One-loop renormalization in a toy model of Ho\v{r}ava-Lifshitz gravity,
    JHEP \textbf{03} 078 (2014)
    [arXiv:1311.6253 [hep-th]].
    
    \bibitem{Barvinsky:2017kob} 
	A.~O.~Barvinsky, D.~Blas, M.~Herrero-Valea, S.~M.~Sibiryakov and C.~F.~Steinwachs,
	Ho\v rava Gravity is Asymptotically Free in $2+1$ Dimensions,
	Phys.\ Rev.\ Lett.\  {\bf 119} 211301 (2017)
	[arXiv:1706.06809 [hep-th]].
	
	\bibitem{Griffin:2017wvh} 
	T.~Griffin, K.~T.~Grosvenor, C.~M.~Melby-Thompson and Z.~Yan,
	Quantization of Ho\v rava gravity in 2+1 dimensions,
	JHEP {\bf 06} 004 (2017) 
	[arXiv:1701.08173 [hep-th]].
	
    \bibitem{Barvinsky:2019rwn}
	A.~O.~Barvinsky, M.~Herrero-Valea and S.~M.~Sibiryakov,
	Towards the renormalization group flow of Ho\v{r}ava gravity in $(3+1)$ dimensions,
	Phys. Rev. D \textbf{100} 026012 (2019)
	[arXiv:1905.03798 [hep-th]].
	
    \bibitem{Barvinsky:2021ubv}
    A.~O.~Barvinsky, A.~V.~Kurov and S.~M.~Sibiryakov,
    Beta functions of (3+1)-dimensional projectable Ho\v{r}ava gravity,
    Phys. Rev. D \textbf{105} 044009 (2022)
    [arXiv:2110.14688 [hep-th]].
    
    \bibitem{Pospelov:2010mp}
    M.~Pospelov and Y.~Shang,
    On Lorentz violation in Horava-Lifshitz type theories,
    Phys. Rev. D \textbf{85} 105001 (2012)
    [arXiv:1010.5249 [hep-th]].
        
	\bibitem{D'Odorico:2015yaa}
	G.~D'Odorico, J.~W.~Goossens and F.~Saueressig,
	Covariant computation of effective actions in Ho\v{r}ava-Lifshitz gravity,
	JHEP {\bf 10} 126 (2015)
	[arXiv:1508.00590 [hep-th]].
		
	\bibitem{Anselmi:2008bq}
	D.~Anselmi,
	Weighted power counting and Lorentz violating gauge theories. I. General properties,
	Annals Phys. \textbf{324} 874 (2009)
	[arXiv:0808.3470 [hep-th]].
	
    \bibitem{Bellorin:2021tkk}
    J.~Bellor\'in and B.~Droguett,
    BFV quantization of the nonprojectable (2+1)-dimensional Ho\v{r}ava theory,
    Phys. Rev. D \textbf{103} 064039 (2021)
    [arXiv:2102.04595 [hep-th]].
    
    \bibitem{Bellorin:2021udn}
    J.~Bellor\'in, C.~B\'orquez and B.~Droguett,
    Quantum Lagrangian of the Ho\v{r}ava theory and its nonlocalities,
    Phys. Rev. D \textbf{105} 024065 (2022)
    [arXiv:2112.10295 [hep-th]].

    \bibitem{Bellorin:2019gsc}
    J.~Bellor\'in and B.~Droguett,
    Quantization of the nonprojectable 2+1D Ho\v{r}ava theory: The second-class constraints,
    Phys. Rev. D \textbf{101} 084061 (2020)
    [arXiv:1912.06749 [hep-th]].

    \bibitem{Faddeev:1969su}
    L.~D.~Faddeev,
    Feynman integral for singular Lagrangians,
    Theor. Math. Phys. \textbf{1} 1 (1969).
    
    \bibitem{Senjanovic:1976br}
	P.~Senjanovic,
	Path Integral Quantization of Field Theories with Second Class Constraints,
	Annals Phys. \textbf{100} 227 (1976)
	[erratum: Annals Phys. \textbf{209} 248 (1991)].
	
	\bibitem{Fradkin:1975cq}
	E.~S.~Fradkin and G.~A.~Vilkovisky,
	Quantization of relativistic systems with constraints,
	Phys. Lett. B \textbf{55} 224 (1975).
	
	\bibitem{Batalin:1977pb}
	I.~A.~Batalin and G.~A.~Vilkovisky,
	Relativistic S Matrix of Dynamical Systems with Boson and Fermion Constraints,
	Phys. Lett. B \textbf{69} 309 (1977).
	
	\bibitem{Fradkin:1977xi}
	E.~S.~Fradkin and T.~E.~Fradkina,
	Quantization of Relativistic Systems with Boson and Fermion First and Second Class Constraints,
	Phys. Lett. B \textbf{72} 343 (1978).
	
    \bibitem{Colombo:2014lta}
    M.~Colombo, A.~E.~Gumrukcuoglu and T.~P.~Sotiriou,
    Ho\v{r}ava gravity with mixed derivative terms,
    Phys. Rev. D \textbf{91} 044021 (2015)
    [arXiv:1410.6360 [hep-th]].
    
\end{thebibliography}
\end{document}